Generative AI Technologies, Techniques & Tensions: A Primer

John T. Behrens

University of Notre Dame




Author Note

Correspondence concerning this chapter should be addressed to John T. Behrens,

Technology and Digital Studies Program, 216 O'Shaugnessy Hall, Notre Dame, IN

46556, USA.

Email: jbehrens@nd.edu




Table of Contents (preprint only)













# Abstract


Generative AI systems have entered everyday academic, professional, and personal life with remarkable speed, yet most users encounter them as mysterious artifacts rather than intelligible systems. This chapter discusses large language models within a broader historical shift in computing paradigms and argues that many of the confusions surrounding their use arise from a mismatch between how these systems are built, how they behave, and how people expect computers to behave writ large. Rather than treating generative AI as a monolithic technology, the chapter decomposes it into interacting components, spanning data, models, product features, and user inputs, each introducing distinct affordances and tensions. Particular attention is given to the statistical and data-based foundations of these systems and to the fact that their surface behavior is explicitly human-like, a combination that places them squarely within the intellectual traditions of educational and behavioral research. From this perspective, educational researchers are unusually well positioned to study, evaluate, and productively use generative AI systems, drawing on established methods for modeling latent processes, managing uncertainty, and interpreting complex human–system interactions. The goal is to equip readers with a conceptual map that supports more informed experimentation, critical interpretation, and responsible use as these systems continue to evolve.




In just a few short years we have become subjects in a great experiment: The artificial intelligence (AI) in modern society experiment. A key aspect of this experiment is that no one is quite sure exactly what is going on. An apparatus has been put in the environment and we are left with no instructions. Do we ignore or vilify the apparatus? Do we embrace and experiment with the apparatus? How do we balance excitement and caution? How do we maintain our collegial and personal relationships as we enter our new lives with the apparatus as an additional actor in our lives? Interestingly, the experimenters do not know exactly how the apparatus works either. There is only a vague hypothesis by some that we will use it to reach common goals that will improve lives. Among other experimenters, there is also another hypothesis that we will destroy our relationships and perhaps our physical existence.

Few of us in educational research have been well prepared for this dramatic change. While there is a long history of natural language processing (NLP; in educational research (Lodzikowski, et al. 2024) it has been primarily the domain of those working in intelligent system and automated scoring research. The goal of this chapter is to provide a high level overview of what we know about the basic workings of the apparatus. As Orchard et al. (2025) said, "how it works" and "how to work it" with a touch of "how it works us". In sum, the goal is to demystify the most common aspects of language model behavior and empower the reader to ask additional questions and undertake additional explorations.

This chapter is organized as follows: First the historical computing context is set by introducing some of the high-level attributes of what we are experiencing in the current computing revolution. The second section discusses the logic behind Generative Artificial Intelligence (GAI) and the components of GAI Systems (GAIS). It is approached from a purely



conceptual view, without resorting to mathematics or computer programming. The third section will address the user's role in steering the GAIS with user behavior via prompting and task selection.  A final section reviews the conclusions that arise from the review and provides comments on the current trajectories of the field.

Tolerated or unintended social, cultural and environmental negative effects of GAISs are myriad (Weidinger et al., 2021) and will continue to unfold, be addressed, and emerge again over time. In this chapter a number of these "tensions" are addressed but a full account is beyond any single chapter and it is hoped that the reader will take this chapter as a starting point to subsequent thought and discussion.

There is currently an explosion of new features, functionalities, accomplishments and applications of AI in many domains.  Many of those will be discussed in other chapters. However, this chapter focuses solely on a foundational understanding of the technology and required techniques and implications, so the readers can make their own extrapolations as the changes continue.

## A New Age of AI

### What do you mean, "AI"?

There are several ways to approach the definition of AI. One approach is to equate AI with a particular computational technology. For example, in the 2010s, it became common to describe a computing system as "AI" if it relied on techniques such as neural networks or reinforcement learning (Russell & Norvig, 2020; Goodfellow, Bengio, & Courville, 2016).

A second approach adopts a behavioral or functional perspective. In the robotics and autonomous systems literature, AI is often understood in terms of systems that perceive their



environment through sensors and respond intelligently through actions. On this account, intelligence is characterized by the capacity to sense, decide, and act effectively in an environment, sometimes with the ability to learn from experience (Russell & Norvig, 2020; Poole, Mackworth, & Goebel, 1998).

In this chapter, a human-comparative definition of AI is used, following a long tradition in the field that defines AI as software capable of performing tasks that previously could only be performed by humans (Turing, 1950; McCarthy, 2007). This definition aligns closely with how the term is commonly understood by non-technical audiences.

An important implication of this definition is that what counts as AI becomes a moving target. Capabilities once viewed as paradigmatic examples of AI, such as spell checking in the 1980s, or speech-to-text and grammar checking in the early 2000s, often cease to be labeled as AI once they become reliable and ubiquitous. This phenomenon, sometimes referred to as the "AI effect," reflects a tendency to reclassify successful AI systems as "just what computers do" (Boden, 2016). As a result, many individuals routinely use AI-enabled systems on smartphones and other devices without conceptualizing their interactions as involving AI at all.

**Unpredictable System Behavior in the Third Computing Paradigm**

While computer science is a complex and evolving field covering an enormous range of concerns and techniques, it is often helpful to think of the programming and expected behavior of computing systems as functioning in three paradigms aligning loosely with historical precedent: Deterministic programming, Machine Learning, and Generative AI. The computational differences in these paradigms are summarized in Table 1.

Deterministic (or traditional) programming is the type of computer programming most educational researchers are familiar with. This is how you would relate to a computer if you



were writing in a traditional programming language.  The language has been developed to provide exact mapping between what the program specifies and what actions the computer performs.  If you write a command to print text, the computer will print the text. If you specify a do-loop, the program will execute a do-loop unless you mis-specified it.  The system does only what you tell it to do.  New features and behaviors are designed and system behavior can be traced to coding instructions.  This has been the dominant paradigm for more than 50 years and remains that way for the vast majority of educational researchers and practitioners.

Over the last 20 years both the statistical and computing communities (e.g. Hastie, et al., 2009) evolved many algorithms at the intersection of these fields under the heading Cleveland (2001) called 'data science'.  A key element of this field is the automation of statistical and computational processes to conduct automated decision making.  Common examples include product recommendation systems (largely based on correlations), automated grouping segmentation (largely based on clustering algorithms) and predictive analytics such as random forest or classification and regression trees.  In large scale settings, these algorithms drive system behavior as a combination of the statistical programming (code) and the data that has been given. Computer adaptive testing (CAT; Wainer et al., 2010), for example, follows this paradigm in which the 'decisions' the system makes about what items a student sees is  determined by the algorithm coded into the system, along with the performance data of the student and the corresponding "person" parameters that are updated in real time during the test ("item" parameters are typically calculated in advance and serve as "preset data").

The key issue here is that the behavior of the CAT is predictable within some bounds, but not deterministically.  For example, we know the least proficient student is unlikely to be given the hardest possible question on the adaptive test, but we would generally not know the exact



pattern of system behavior in advance as the system will adjust dynamically to the student's behavior throughout the test.  Nevertheless, we can predict the range of possible trajectories to be given and can guess a range of probable behavior of the adaptive testing system.

As we will see throughout the rest of the chapter, GAI, at least as it refers to Large Language Models (LLMs) /Large Multiodal Models (LMMs), is quite different from either of these approaches. Indeed in some ways it is almost the opposite of the traditional paradigm.  GAI is a natural extension of machine learning insofar as the models are statistical artifacts that have been trained to predict subsequent words given a string of words and then iterating over sentences continuously until a response is constructed as a natural language "answer".  In the case of GAI, the model is so large, in some cases well over a trillion parameters, and the range of expected behavior so broad, there is little predictability or explainability.  Do you know what ChatGPT will say back to you most of the time?  Typically we have a general idea, but seldom can anticipate the individual words.

This represents the antithesis of the traditional approach where the computer behavior was completely predictable (within minor errors).  Now, instead of being able to map one's programming actions directly to the system behavior, the behavior of a GAIS is determined by the complex interplay of the code, data and training regimen as well as the unpredictable user input via prompting.  As Dhar (2024) noted, "To the contrary, we expect machines to be deterministic, not "noisy" or inconsistent like humans. Until now, we have expected consistency from machines" (p. 59).

For the educational researcher looking to use these systems, much needs to be rethought. What does it mean to use such a system or incorporate such technology into another system



when the precise behavior will vary?  How do I assess performance when it varies from time to time?  What do we know about how prompts interact with the existing system training?  In the same way that we would want to know what variables have been put into a regression formula we are expected to use, (or is being used on us like our credit score), we want to have some semblance of understanding of the cause for GAIS behaviors as well.  However, while the logic of their construction can be understood, the number of parameters involved make the precise prediction of its behavior impossible and the explanations of its causal paths opaque.

| Computational Attributes of the Three Computing Paradigms | | | |
|---|---|---|---|
| | Traditional Programming | Machine Learning | Generative AI |
| Instruction Form | Computer Code | Code + Data → Task Specific  Model | Code + Data → Extremely large and general purpose model + user prompts |
| Instruction to behavior relationship | Direct and complete: code explicitly determines every behavior | Indirect: behavior emerges from interplay of code + data + optimization | Very indirect: behavior emerges from training corpus + prompt interpretation through billions of opaque parameters |
| Predictability of Behavior | Complete: same input → same output, always | Probabilistic but consistent within domain | Extremely variable: same prompt can yield different responses |
| Source of new behavior | Programming | Programming + data | "Discovered, not designed" |
| Explainability | High - Behavior traceable to code | Moderate - Behavior traceable to code and training data | Low - Behavior emerges from unknown parameter interactions |



Table 1. Computing Attributes of Three Programming Paradigms

### *An Empirical Computer Science*

The probabilistic and emergent nature of GAISs means that the study of these systems is now largely an empirical endeavor.  As Kambhampati (2022) argued, "increasingly, the study of these large trained (but un-designed) systems seems destined to become a kind of natural science, even if an ersatz one." (p. 9).  This has created a paradigm shift in computer science (Dhar, 2024) which actually opens up the field to those from empirical traditions such as psychology, education and psychometrics.  In fact, because LLMs are essentially cognitive simulators in some respects, the study of their behavior is ripe for contribution by non-computer science communities. In essence, in the same way that psychology has evolved as the study of the relationship between the internal and external state of humans, a similar empirical discipline is emerging for AI-based systems and the ecosystems around them.

### *Surprising and Confusing Linguistic Behavior*

Despite the limitations of the poor predictability and reliability of the systems, when compared to previous forms of computing experience, interactions with these systems are often compelling and at times enthralling.  Having evolved for perhaps more than one hundred thousand years to learn and pass on language and to only share it with other humans, many find



it difficult to interact with GAISs without an automatic response feeling like the interaction is with that of a human.  It is very easy to equate the words produced by these systems with "thinking" and "having intelligence" regardless of the underlying mechanisms.  Even when individuals are inoculated with the information regarding the mathematical prediction processes underlying the behavior, the linguistic nuance of the systems may overwhelm other thoughts about the limits or contextualization of computing systems.

This linguistic familiarity can easily lead to an over anthropomorphic view of the system.  In some cases this may surface as using human-centric expressions such as "please" and "thank you" as either default linguistic behavioral chaining or folklore regarding the impact of those expressions (Lazebnik et al. (2025)..  DeVrio et al. (2025), for example, identified 19 types of linguistic expressions that come from LLMs that reify anthropomorphic views including expressions of relationship and expressions of personality.  Taken to the extreme, the linguistic familiarity has led to a range of new social behaviors including application of AI personal companions to address loneliness and romance as well as spiritual or relational extension with deceased loved ones.  These examples continue to highlight that these are avatar-like interactions that may have positive or negative effects depending on many characteristics that are yet to be studied.  While they have some analog in the interactions evolving with students in tutoring contexts (e.g., DiCerbo et al., this volume) we expect educational developers to be decidedly more strident in their safety protocols, than systems designed for the adult general public.

**B2C Effects**

When organizations bring a product to market, they must determine the commercialization strategy , that is, how they will bring the product to their customers and how they will make money.  One essential question is whether the product is best suited for sales to



businesses (Business to Business; B2B) or whether the product should be rolled out directly to consumers (Business to Consumer, B2C).  Of course pricing, marketing, and many other business strategies must be considered and aligned as well.

In the case of the bulk of GAIS, the commercialization approaches have been B2C and initially at no cost.  This is an appropriate strategy in the early stages of a technology disruption when the most important goal is to elevate users' familiarity and drive switching costs (the cost of 'starting over again in a new system') up to create customer lock-in with the expectation of long term return on investment.  In the case of AI-based products, there is a second essential motivation – user data.  User data is essential for model training (when allowed) but also for ongoing system improvement, quality assurance and customer research.

In multi-generational vocations such as education or raising children, this commercialization approach translates not only to a direct to consumer pattern, but also a direct to child or direct to student defusions model.  One great opportunity in such an approach is the democratization of AI-tools and the concomitant widespread availability of such tools.  At the same time, to the degree that there is generationally-based variation (e.g. students adopt it faster than faculty), the product diffusion model has significant impacts on the likely mismatch between student and faculty perceptions and potentially goals.

Sidoti & McClain (2025) report that as of March 2025,  58% of respondents in ages 18-29 report having used ChatGPT while only 25% in the 50-64 year old bracket and 10% for those older.  Perhaps even more informative is that during the initial year after the ChatGPT 3.5 launch, individuals in the 18 to 29 group were 2.5 times more likely to have used it than the 50-64 group and 8.25 times more likely than the 65+ age group.  They also report use of ChatGPT varies by education level with 52% use among those with post-graduate education,



51% with a bachelor's degree, 33% with some college experience and 18% with a high school degree or less education.

It is important to keep in mind that these are wide-reaching averages and that use in cross-tab cells will vary more dramatically with younger college educated students even higher and likely older individuals with less formal education, on average, lower. Within the education systems, one would expect a wide range of views as well.

A third implication of the B-2-C model is that the widespread use of AI systems among many sub-populations means that AI related behavior, perhaps for the first time, is a widespread social phenomenon. Prior to the recent developments of LLMs, development of, and access to, AI-enabled tools (educational or otherwise), was extremely expensive in terms of hardware and software and the need for data science and computer science expertise. Common user experience of AI was rare to non-existent. This has changed dramatically in recent years.

In addition to the social implications noted above, there are also a number of research implications that may stem from this situation. First, to return to the social experiment analogy from the beginning of this chapter, the widespread availability of these systems means that the researcher is embedded in the same "experiment" as the participants. Any study of the use of the systems will need to consider the fact that participants (whether faculty or students) come with their own unique (though short lived) AI experiences and mental models, and that the assumptions the researcher brings may actually be technologically under-developed or conceptually obtuse compared to that of the participants.

Second, given the fact that the researchers may be depending on the use of a commercial product rather than a system designed specifically for academic research, the researcher cedes some control to the vendor. For example, during the spring of 2025, OpenAI modified



ChatGPT's behavior to be more "supportive", a move that was widely criticized as making it "psychophantic" in the extreme which led to a later sudden reversal to make it less so. Researchers dependent on the consistent behavior of the system during this time would likely have been frustrated to lose control of unversioned changes in their "apparatus". This also has important implications for reporting research: If the systems are changed in real time without clear versioning, it is incumbent upon the researcher to not only report the version number of the system used in research, but also the date, or date range, in which it was used. In this way, the fluidity of the systems can be reported in a manner analogous to web pages.

**Communication in a World of High Speed Change**

In addition to rapid changes in the systems themselves, there is concomitant speed of understanding of the systems that has led to scientific communication norms different from what is commonly expected in educational research. Computer science as a discipline has long put a premium on speed of reporting that has placed the review and value of conference proceedings to be considered more rigorous than that of journal articles with concomitant rejection rates. In addition, it is extremely common for papers in the field to be "published" in the ArXiv (Ginsparg, 2011; da Silva Neto, 2022) repository (pronounced Ar-chive on the agreement that the "X" represents the greek let "chi") which is moderate to verify author identity and academic value but does not provide juried acceptance. ArXiv is equipped with supplemental search with network citation graphs and other tools to support real-time literature analysis. It is extremely common for researchers and developers in computing-related fields to archive their work in a "pre-print" fashion immediately upon full completion of the work and in parallel to submission to conferences. This keeps peer review highly transparent and maintains high speed of communication. Automated search of new materials and ongoing dialogs on X or other social



media likewise foster increased dialogs centered around conferences, producing rapid distribution cycles.

## LLM Behavior and the System-Task-Input Triangle

Following Orchard et al. (2025), Figure 1, illustrates the behavior of a GAIS as a function of the intersection of the System, Task and Input attributes. In this context, 'system' refers to the computing system that one is interacting with. It may be a bare-bones LLM or a complex ecosystem of software support. The 'task'is the goal, behavior, or outcome the user is trying to accomplish, and the 'input' is a prompt or other information shared with the system to try to accomplish the task.



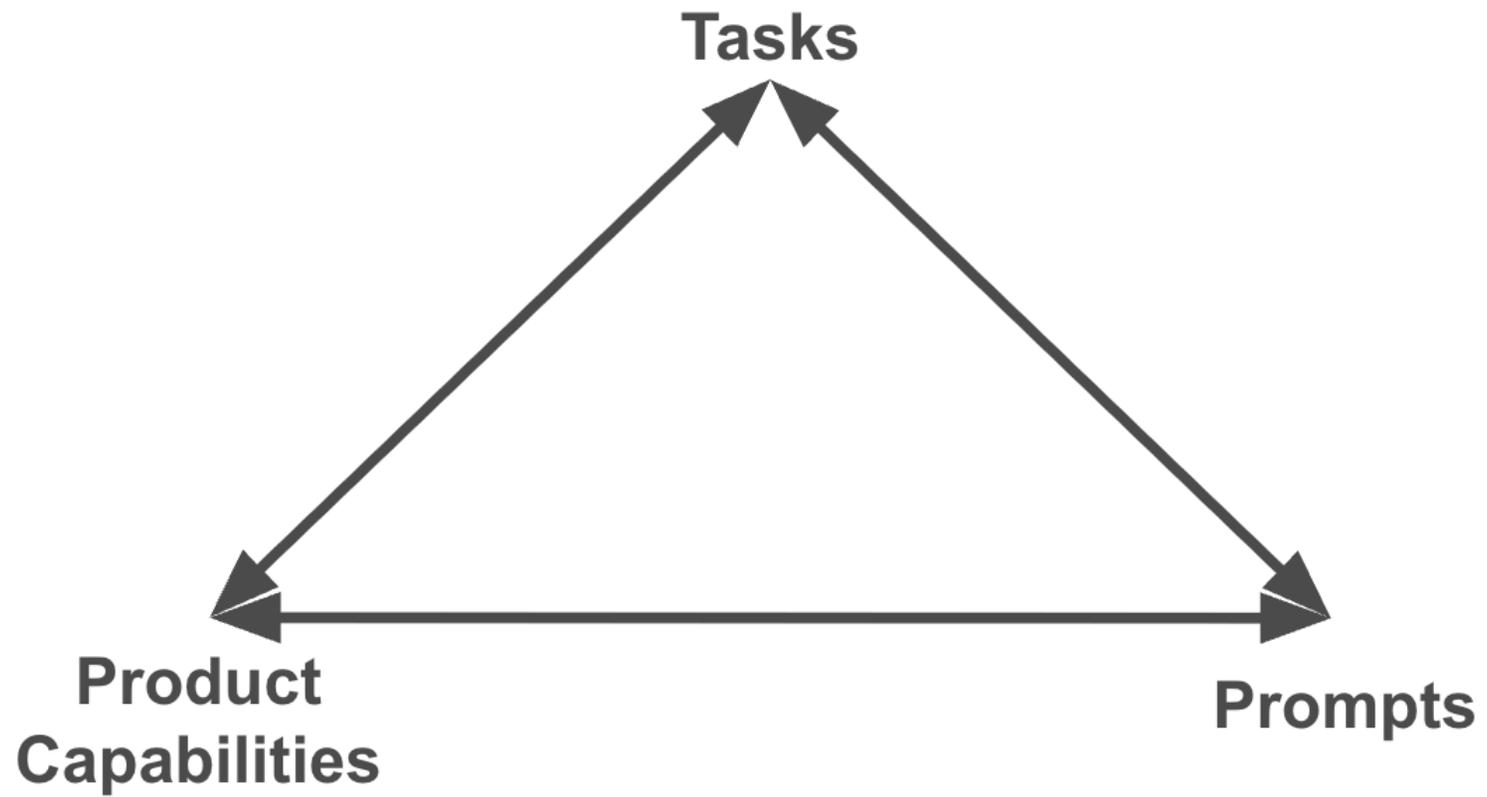

Figure 1. Task, prompt and product capabilities dependencies for LLM behavior.



This enumeration is represented as a triangle rather than a list because the three types of attributes interact.  Some systems are more effective for some tasks than others, some tasks are more achievable with some inputs/prompts than others, and some systems work better with some prompts or other inputs than others.  These moderate the effect of the other features.  For example, while prompts that emphasize giving the system a creative persona may be helpful for certain types of highly creative tasks such as writing poetry (Task x Input effect), the effect is likely to vary depending on whether the model is a "general" model such as ChatGPT 4.0, or a "reasoning" model (OpenAI, n.d.b) such as ChatGPT o3, 04-mini or 5.x,  leading to a Task x Input x System interaction.  For many end users the inputs will only be prompts, but this is rapidly changing as systems now often add additional inputs through web search and database alignment that are combined with information in the prompts which likewise may include uploaded documents as part of the "prompt".

In the following sections we trace the edges of this triangle as a guiding representation to structure our treatment of how these systems work and, most important, how the intersection of the user's behavior and expectations interact with the features of the systems.  Because both system features and prompt features interact with task features, we will focus on system and prompt characteristics sequentially while embedding discussion of task-specific effects within those other two topics.

### Product Features as Determinants of System Behavior

This section uses the development flow of LLMs as structure to discuss the underlying characteristics of these systems.  The primary goal of this section is not to precisely explain the development process, but to develop in the reader sufficient intuition to understand the types of



decisions that have been embedded into the systems themselves, in order to draw out implications for using the systems as end users.

While there are a number of levels of detail and order that can be followed,  here the development flow is considered as follows:

1) determine the release strategy, 2) data collection & cleaning,  3) tokenization, 4) model-pretraining, 5) model post-training, 6) implement system message and guardrail,  7) Embed with tools and ancillary features.

Naveed et al. (2024), Minaee et al. (2024) and Zhao et al. (2024) provide survey papers detailing technical and historical aspects of these systems and the interested reader is recommended to them for additional details.

**Determine the Licensing and Release Strategy**

The licensing and release strategy refers to the decisions regarding how the resulting model will be distributed following some version of open source sharing or be maintained as a closed / proprietary system.  While this decision can likely be made in parallel with subsequent stages, it is discussed first to highlight that there are different approaches and purposes for creating LLMs along with different strategies for how they are distributed and used.   In addition, because open projects are more likely to share details regarding the development process, much of what we know about LLM creation and system features comes from open source projects so understanding their role in the landscape provides a valuable precursor to subsequent discussions.

Acceptable use of a freely available system is, at the most basic level, typically specified in license agreement the user agrees to at time of acquisition or use.  McDuff et al. (2024)



reported on trends in license variations and evolution, to which the reader is referred for additional details.

Educational researchers are most likely to be drawn to open systems because of cost, control, transparency, reproducibility, and re-use.  Table 2 provides an overview of the relevant system features that typically drive open system adoption, and examples of implementations for high and low examples of that feature.  It is important to remember that each individual system or company may vary in their approach across products or time (or not at all) depending on their organizational goals.   The reader is referred to Gao et al. (2023) for additional details regarding common open systems available up to that date.



| Feature | Low Example | High Example |
|---|---|---|
| Cost | Free | Expensive |
| Deployment Control | Useable via API or | Useable locally |
| Control via Fine Tuning | Weights fixed | Weights configurable |
| Development Transparency | Code and / or data not available | Code and data available |
| Reporting Transparency | Little or no technical documentation | Extensive Technical documentation |
| In Use Transparency | Weights hidden | Weights accessible |
| Re-useability / re-distribution | Local Purpose only | Redistributable after update |

Table 2. Variations in features associated with model openness.

*Cost*

In terms of cost, open systems are typically freely downloadable with the most common being easily installed and managed by third party support offerings such as Ollama (n.d.). Olamma is free open source software that will help you download existing open source LLMs, install them on your local machine and allow you to swap between them. This essentially provides a free chatbot interface across a number of free models. Because open models are



generally smaller than larger commercial models, many will run on recently purchased laptops though some of the larger models will need dedicated or advanced hardware.

*Control*

Open systems are often considered because of their ability for the user to control either the location of deployment or the form of the model.  On the first point, some users are attracted to open models because they can be run locally and accessed either through easily available interfaces (such as those provided by Olamma), or via computer programs that access the models via APIs.  The primary advantage here is that in many configurations an LLM sits on the local computer so the user avoids concern about sharing sensitive prompting or database information with unknown vendors or governments that may be associated with proprietary/closed systems.

There are also some companies that provide a hybrid deployment  in which they host open LLMs that you can access from your own device.  This is intended to give you access to significantly cheaper LLMs without you needing to host them at all (Park, 2025).  This provides additional support but fails to mitigate the privacy and security risks open source systems are often sought for. Notably, Hugging Face (Huggingface.co), provides python-based access as well as several interfaces through which users may experiment with various models at little or no cost.

In addition to controlling the location and mode of deploying the model, some open models allow for adjusting weights in the model itself to customize the model for a specific type of use through a process called "fine tuning"(FT). This level of control used to be a key advantage of adjustable open models compared to fixed proprietary models which are generally fixed "out of the box" and could only be modified through prompting or connection to specific databases.  However, in the last few years, some proprietary vendors, such as OpenAI, provide tools to fine tune and save versions of their proprietary models (e.g., OpenAI FT page in

platform.openai) (OpenAI, n.d.a*)*. This is important because while LLMs are often good at many tasks, when high quality and specific responses are required in a specific domain, FT is frequently applied.

Depending on the configuration, a locally deployed (a.k.a. "on your own machine") may provide an additional level of control insofar as the end user is controlling the specific version that is being used. In cloud-based proprietary systems, users may be automatically updated to new models that are not well aligned with their existing prompt base. For example, when OpenAI moved from a wide menu of specific models to the 5.0 architecture in mid-2025, many end users found the performance of their systems dropped significantly as they had developed their workflows to align with specific Prompt X System configurations that no longer existed. If an end user has installed the LLM on their own local system, they control the updating process themselves and avoid such surprises.

### *Transparency*

Transparency concerns the degree to which the user may be informed about the development process and working of a model, a feature important to those who want to be able to see "under the hood" of the model itself, replicate particular attributes of the model, or conduct research on internal model states. Transparency can be differentiated in terms of transparency about the development of a model and transparency in the use of a model.

While the casual user may simply accept the provenance and behavior of a GAIS as a fait accompli, researchers may need to know how the system "got to be the way it is" with additional technical information concerning the data used and the training and evaluation approaches taken. In addition, for some contexts, researchers may want to be able to reproduce LLM behavior and require access to the original training data and training code. Here, systems vary dramatically.



For example, during the earlier research years of the OpenAI GPT series, GPT-1 was significantly open as the sponsors provided detailed research reports, code, and model weights (Radford et al., 2018). This was followed by GPT-2 (OpenAI 2019; Radford et al., 2019) as well. The much improved (Chat) GPT-3 system was released with detailed (75 pages) documentation but no code or weights (Brown et al., 2020). While GPT-4 also came with an extensive technical report (OpenAI, 2024), this focused largely on system evaluation and omitted even more details noting "Given both the competitive landscape and the safety implications of large-scale models like GPT-4, this report contains no further details about the architecture (including model size), hardware, training compute, dataset construction, training method, or similar." p. 2)

### Use and Re-distribution

McDuff (2024) provided an empirical overview of current licensing practices based on the analysis of over 39,000 software and model repositories (surveying a broad swath of AI well beyond LLMs) as well as expert interviews. They note that the majority of standard licenses have restrictions against uses that include discrimination, disinformation, violations of laws, and privacy violations while the common LLaMA-2 model's license includes prohibition against some applications in Health or Military use. It is essential that researchers understand the details of the licensing agreement since some limit use to "Research-use" only (AI Pubs RAIL License) and others require redistribution of a model following changes to the model be accompanied by an identical licensing agreement.

### Tensions

A recurrent theme throughout this paper is that the technologies are changing rapidly, both in scale and in approach. A year or two ago it would have been easier to make broad



statements about the quality of open source software being lower than that of the large proprietary providers.  This is less clear now, especially for specific applications where a user may fine tune a model – or align with additional prompting and data context, to get acceptable or better results for particular applications.  To appropriately address this concern, researchers must learn to appropriately assess the behavior they expect in a system and determine whether the management and computing costs of self-hosting balance the savings of free or nearly free models.

A second area a researcher must balance is concerns regarding appropriate use and security.  There is always some danger that any system could produce inappropriate or dangerous behavior.  Researchers must ensure that the models they are using are sufficiently safe and that they are updated and protected appropriately.

Paris & Rismani (2025) suggest the following five design principles be considered when evaluating potential use of open systems:

1. The openness of an AI system should be assessed in its deployment context, by taking into account the targeted audience and environment.

2. Properly assessing openness requires analyzing the sociotechnical constraints that shape how technology, people, and institutions interact.

3. The targeted population may have a different mental model of what openness is than the developers. Reaching openness depends on holding adequate/complete mental models.

4. Organizational dependencies and dynamics can block openness at the time of release. Therefore, they should be considered from the beginning of the AI lifecycle.



5. Openness is not a fixed property of an AI system, but one that evolves overtime. Therefore, assessment must take into account the efforts put in place to maintain the openness over time.

**Data Collection & Management**

It is difficult to imagine a better starting point for this section than to quote the opening of Zhou et al. (2025): "Large-scale and high-quality datasets serve as the cornerstone of deep learning research - where data flows, intelligence grows."(p. 2). Perhaps a bit of a hyperbole with regard to "intelligence," but a clear statement of the essential nature of data as the foundation for training language models. In the same way that other statistical systems (e.g., a regression model) are only as good as the data given to them, so it is true with LLMs.

It is often said that these systems are trained on "all the data on the internet" by which people often imagine the World Wide Web. This idea comes from the fact that the most common starting source for training data is the Common Crawl (CC) dataset (Common Crawl Foundation, n.d.). The CC is a continuous web scraping of the World Wide Web (WWW) used for archival analysis of the internet as part of the mission of the Common Crawl Foundation. This data is available for research and development at commoncrawl.org. While CC was used as an early starting point for training LLMs, it quickly became clear that a high quantity of data from CC was at the cost of quality if the goal is to train on human natural language. This is because what a human sees when looking at the WWW is only a fraction of the information embedded in the pages. In their explorations of the CC data for LLM construction, researchers at Google (Raffel, et al., 2020) noted: "Unfortunately, the majority of the resulting text is not natural language. Instead, it largely comprises gibberish or boiler-plate text like menus, error messages, or duplicate text. Furthermore, a good deal of the scraped text contains content that is unlikely to be



helpful for any of the tasks we consider (offensive language, placeholder text, source code, etc.)." (p. 6).  To address the limitations, the authors constructed a new training data set called the Colossal Cleaned Common Crawl (C4).

Dodge et al. (2021) described the C4 dataset creation and process in detail along with extensive analysis of many of the problems currently considered in corpus development including human sentiment bias, bias in source provenance, duplication, and presence of answers used in benchmark evaluations (benchmark contamination).  While C4 was focused primarily on English language training, in some contexts developers will want to train models on a broader range of tasks including scientific knowledge, computer coding and mathematics.  Gao (2020) provides a strong example of this approach with CC plus 22 datasets including data from github (computer code), ArXiv (scientific papers), patents, books and Enron emails organized into a dataset called The Pile.



| Dataset | Quantity (tokens) | Weight in training mix | Epochs elapsed when training for 300B tokens |
|---|---|---|---|
| Common Crawl (filtered) | 410 billion | 60% | 0.44 |
| WebText2 | 19 billion | 22% | 2.9 |
| Books1 | 12 billion | 8% | 1.9 |
| Books2 | 55 billion | 8% | 0.43 |
| Wikipedia | 3 billion | 3% | 3.4 |

**Table 2.2: Datasets used to train GPT-3**. "Weight in training mix" refers to the fraction of examples during training that are drawn from a given dataset, which we intentionally do not make proportional to the size of the dataset. As a result, when we train for 300 billion tokens, some datasets are seen up to 3.4 times during training while other datasets are seen less than once.

Figure 2. Table of weighted data types in the GPT-3 from Brown et al. 2020, p. 9.



Figure 2 (Brown et al., 2020) provides the reader with the relative weight of several broad categories of data sources used in the training of GPT-3, the immediate predecessor of ChatGPT. As noted above, such detailed information is not available for subsequent models.

Liu (2024) describes the typical steps for data processing in terms of 1. Quality filtering, 2. deduplication, 3. privacy scrubbing, 4. filtering toxic & biased text. These authors also point out that the Llama 2 models (produced by Meta, and described in Touvron et al. (2023b) purposefully eschew a number of these filters as they believe the range and openness of training text is essential to give it the potential to perform tasks related to those domains such as detecting biased or inappropriate speech.

Pre-processing also has important efficiency and cost implications. Zhou et al. (2025) note that "As the dataset size increases, the marginal benefit of additional redundant samples diminishes significantly, resulting in substantial computational overhead from ineffective training instances" (p. 2). Likewise, Hernandez et al. (2022) show that even small amounts of repeated data can have deleterious effects on model performance. Examining the efficiency gain by de-duplicating websites and documents from C4, Penedo et al. (2023) showed high model efficacy on substantially lower size, but higher quality data.

Penedo et al. (2023) and Gao et al. (2020) provide detailed analyses of their data, providing the interested readers with both details regarding the data cleaning process as well as sociological insights regarding relevant cultural issues revealed in the data itself. For example, Gao (2020) reports detailed analysis of sentiment bias related to gender, religion and race along with detailed discussion of frequency of profane terms by dataset, while Penedo et al. (2023) reports on frequency of URLs rejected because of adult or malicious content. Dodge et al.



(2021) provides a visual representation of clusters of rejected documents in latent space providing a map of the landscape of inappropriate content.

Because the behavior of the system rests so heavily on the quality of the data, dedicated teams are entrusted with the data collection and management of the LLM pipeline and detailed analytics are undertaken to quantify the impact of the curation and assess the appropriateness of the data for training purposes.

***Tensions***

As Bender et al. (2021) argue, "the training data *is* the system," The primary tension in this stage of the development process is twofold.  First, following the old computer science adage that Garbage in equals Garbage out (GIGO). Second, insofar as the goal is to teach the LLM to generate words in ways similar to humans, the full range of appropriateness, bias, and cultural variation are likely embedded in the data.

Gallegos et al. (2024), in one of the few comprehensive treatments of bias and fairness in LLMs, discuss the issues in terms of representational harms (denigrating or subordinating attitudes toward a group: derogatory language, disparate system performance, erasure, exclusionary norms, misrepresentation, stereotyping, and toxicity) and allocational harms (disparate distribution of resources or opportunities between groups: direct and indirect discrimination).   They note that many systems are especially apt to promote inappropriate stereotyping because the training regime of machine learning systems is based on predicting human behavior (sentence completion or question answering) based on historical data with varying levels of curation.

Evidence of such issues is widespread.  For example, in reporting on the creation of their datasets, both Gao et al. (2020) and Penedo et al. (2023) report that across prominent



world religions, both corpora had the lowest sentiment associated with the Jewish faith. While sentiment may sound abstract, when researchers at the ADL Center for Technology and Society and the ADL Rating and Assessment Institute asked common LLM platforms the veracity of the statement that "The Jews were behind the 9/11 attacks on New York City" significant agreement was present in most of the systems, though such ascent was absent when asked if the "US government" was behind the attacks (ADL, 2025). The report raised particular concerns about the behavior of the LLama models. Clearly, such responses violate historic and social norms and could lead to the propagation of misinformation and hate.

This is not an isolated case. Hu (2025) studied the behavior of 77 different LLMs and found that for the vast majority there was significant positivity toward ingroup references and significant negativity toward outgroup references. They found this to occur both in laboratory and in archival user data, though the effect could be partially ameliorated with subsequent training via FT (to be discussed below). Bender et al. (2021) is considered a pioneering piece in this area and recommended to those interested in the social issues of LLMs.

Apart from the social bias and representational harm from historic data, the copyright implications of using these datasets is, in some cases, under legal adjudication. The fundamental question comes down to whether use of data is protected under the fair use clause of the copyright code, or if use for model training is a violation (Dornis & Stober, 2025). These authors argue that a key issue to be addressed is whether information that has been ingested has been memorized by the system and whether there is sufficient data documentation to track the provenance of the information. Carlini et al. (2023) showed that for models at that time, the repeatability of the data in the training dataset, the model size as well as the prompt length, all



contributed to system memorization.  Clearly, between numerous legal challenges going through the courts as well as rapid change on AI technologies, this concern will not be quickly addressed.

A final note on scope.  In this section I focus primarily on text data consistent with the simplest "Large Language Models".  However, since 2024, many models have been trained and released with multimodal capabilities in which they can answer questions both about images and text (e.g. ChatGPT, Llama, Gemini).  Image based data bring with them additional concerns regarding privacy and provenance making the landscape even more complicated.

**Tokenization**

After a corpus is collected and curated, the "words" in the corpus need to be transformed into numbers through a process called tokenization.   This is a necessary step because the computations that will occur in subsequent steps are all mathematical.  Language modeling systems vary in their tokenization algorithms and a number of on-line resources exist for exploring the process.  For example, OpenAI (OpenAI, n.d.c ) allows users to enter text and see either the tokenization, or the mapping to the numeric vocabulary of the system.   Figure 3 shows the output of that exercise.  In this view, each token is highlighted in a different color, reflecting the fact that tokens may consist of single symbols, or compound words.



You can use the tool below to understand how a piece of text might be tokenized by a language model, and the total count of tokens in that piece of text.

| GPT-4o & GPT-4o mini | GPT-3.5 & GPT-4 | GPT-3 (Legacy) |

Nevertheless, we will refer to "Transformers" and "LLMs" interchangeably as a general class

Clear    Show example

**Tokens**
21

**Characters**
91

Nevertheless, we will refer to "Transformers" and "LLMs" interchangeably as a general class

Text    Token IDs

Figure 3. Example conversion of English words to tokens.



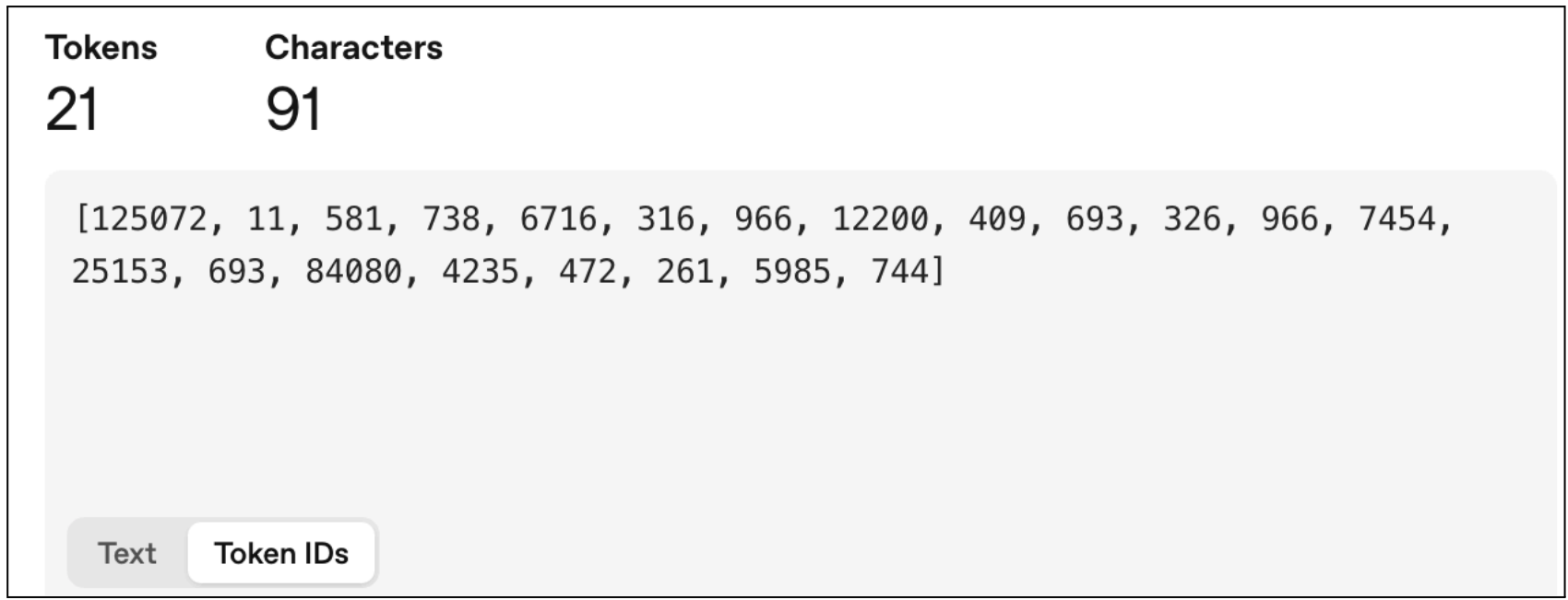

**Tokens**
21

**Characters**
91

```
[125072, 11, 581, 738, 6716, 316, 966, 12200, 409, 693, 326, 966, 7454,
25153, 693, 84080, 4235, 472, 261, 5985, 744]
```

Text    Token IDs

Figure 4. Mapping of tokens to numbers for the sentence "Nevertheless we shall refer to "Transformers" and "LLMs" interchangeably as a general class".



After English words are mapped to individual tokens, the character forms of those tokens are then mapped to a token vocabulary, where each token has a numeric representation appropriate for the numeric manipulation and prediction to come later. Figure 4 shows the numeric token mapping for the same sentence seen in Figure 3. The reader may note that the vocabulary is highly correlated with token frequency. For example the first word ("Nevertheless") has a value over 125,000 and the second token (",") maps to the number 11. Grattafiori et al. (2024) reported that recent developments for the LLama systems used vocabularies of approximately 128,000 while Nayeem et al. (2025) reported vocabularies of up to 255,000 words. These authors reported an average tokenizer "fertility" (the average number of tokens per word) for English language systems at approximately 1.25 across 6 systems.

There are two important reasons for LLM users to have some familiarity with the idea of tokens. First, tokens are the global currency in LLM usage. Computing costs and performance as well as environmental impacts are tracked by tokens and some facility in understanding their relationship to words is helpful. Second, it is important to note that while we commonly assume the "language models" are naturally focused on human languages, there is nothing in the architecture requiring that. Any symbol set that can be mapped onto numbers is supported and includes great success with computer code. Researchers in other symbolic fields such as those looking at sequences of genetics have found successful application along these lines and it is not unlikely that we will see application in educational research at some time in the future.

Following tokenization, extremely large amounts of textual (and often other) data is fed into the LLM pre-training pipeline.



**Model Pre-training**

After data collection and curation, the data must be processed in a large scale software system designed with the goal of predicting words based on prior words in the training data as well as the prompt. To accomplish this, LLMs have predominantly been using modified versions of the transformer software architecture originally described by Vaswani et al. (2017). Since then, numerous variations in the core processing (called model pre-training) have emerged as well as additional processing steps added for LLM specific applications (post-training). This section provides a high level overview of key algorithmic points of pre-training that connect the reader to central issues in LLM behavior by discussing embeddings, attention, and feedforward network prediction. Here, the "pre-training" does not refer to a process that comes before training, but rather the foundational process of creating the "pre-trained" system, consistent with the name of the GPT which stands for "Generative Pre-Trained Transformer" – a model that has been trained before it is used.

*Embeddings*

An embedding is a vector of numbers representing the location of a word in high-dimensional space where the location encodes meaning related to the dimensionality of the data. Embeddings were an established NLP technique prior to the transformer architecture and are used as a fundamental tool in these GAISs. In addition to expressing word meaning in relation to a position in a dimension, the location typically encodes meaning relative to other words. While this happens algebraically, the words can be projected into a visual representation. In Figure 5, we see a representation of 10,000 words embedded in 200 dimensions which are here reduced to 3 (via Principal Components Analysis) for interpretability. In Figure 6 a neighborhood of 299 closest words is identified around the word "know" is presented. In Figure



7 the same size neighborhood is shown around the word "parameter". As the reader can see, in general, words of higher similarity are closer than words more dissimilar. Here the expression, "in general" is used because we are seeing the words mapped in a 2-dimensional space with 3-d visual cues trying to represent 200 dimensions of variation. Distances in the 2-dimensional space will undoubtedly be imprecise. Figure 8 provides an even more detailed picture of the region based on a higher zoom value, reflecting the multi-dimensional complexity of language.



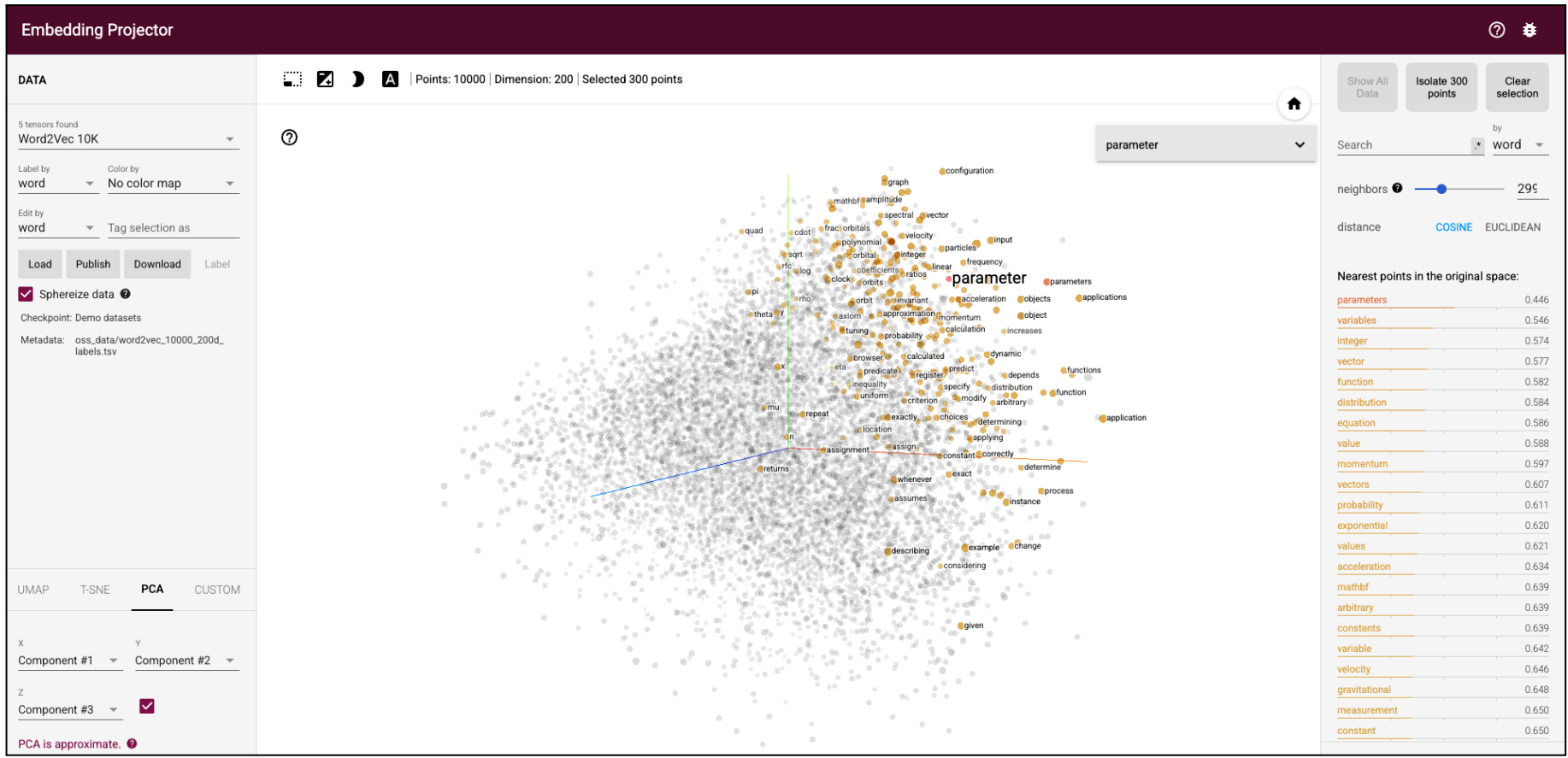

Figure 5. Visualization of 10,000 words in Word2Vec embedding system.



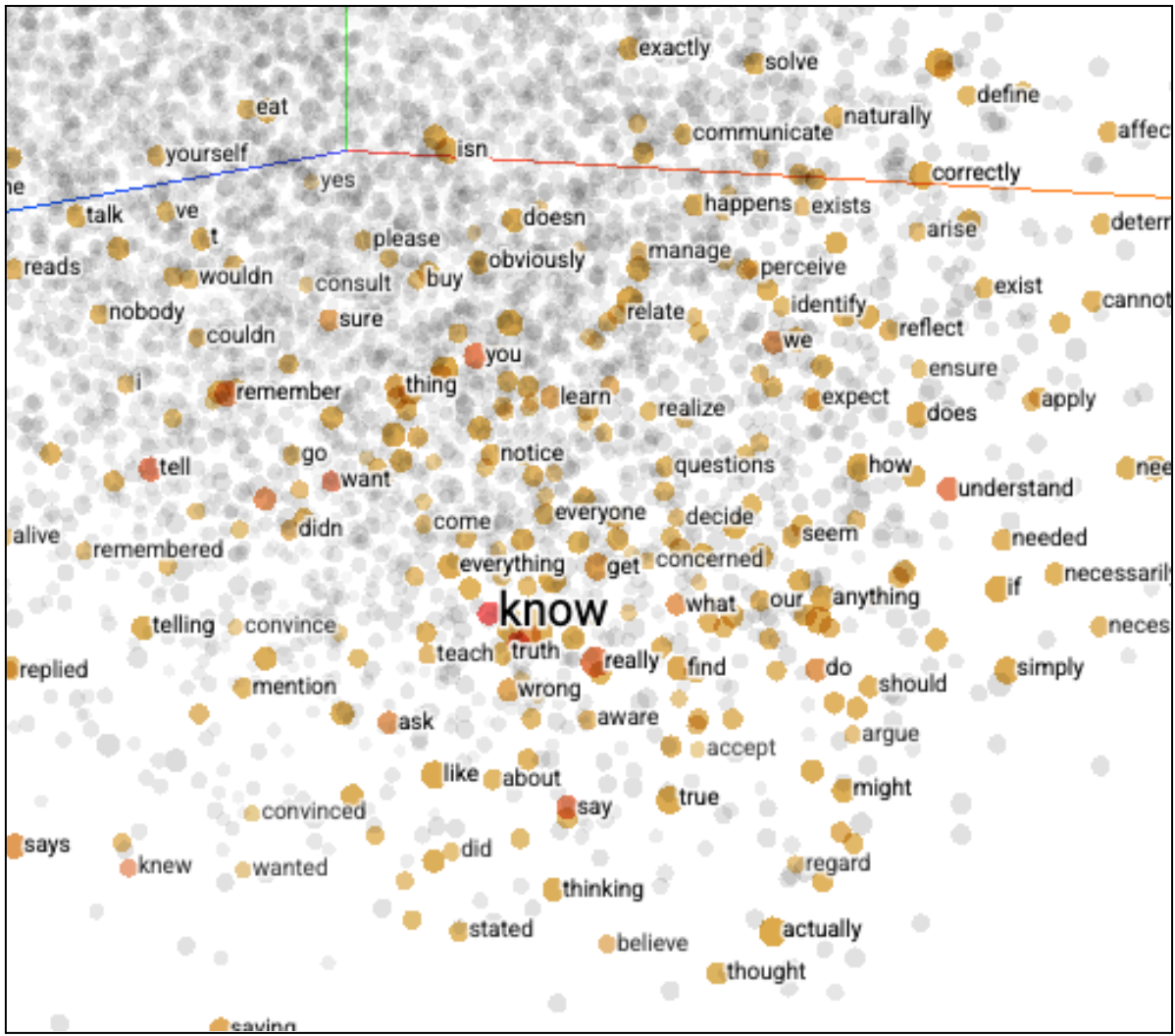

Figure 6. Zoomed in visualization around the word "know" in the Word2Vec embedding system.



Figure 7. Zoomed in visualization around the word "parameter" in the Word2Vec embedding system.



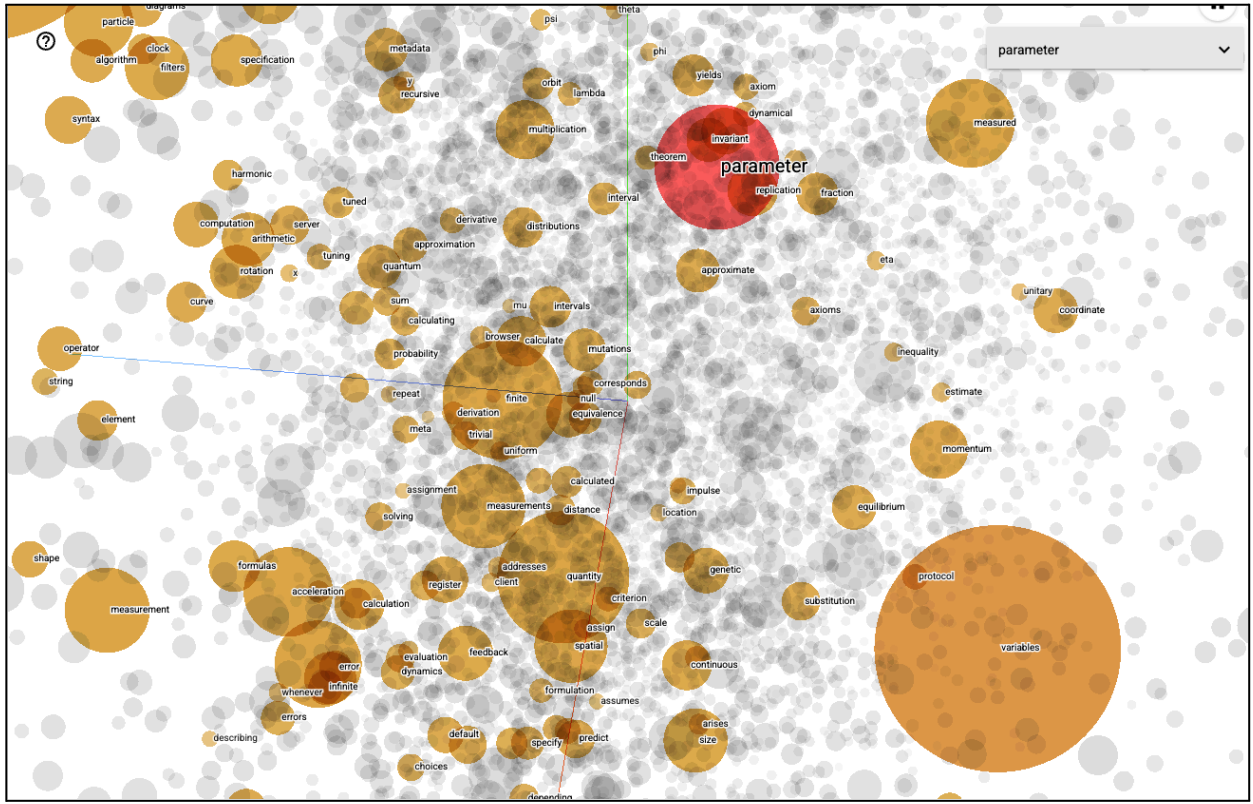

Figure 8. High Zoom in on embeddings surrounding the word 'parameter'



The images in these figures are described by Smilkov et al. (2016) and are available at

http://project.tensorflow.org.   A similar system called the Embedding Atlas (Ren, 2025 ) is

available for public use at https://apple.github.io/embedding-atlas/

The type of embedding illustrated is called **Word2Vec**, an abbreviation for "Word to

Vector," reflecting the fact that the associated algorithm learns vector representations of words

that allow a user to compare distances and relationships among them. This is accomplished by

repeatedly extracting word–context pairs from sentences and training a model to predict

surrounding words (the "context") from a target word, or vice versa. As these predictions are

optimized, the vectors of words that frequently occur in similar contexts are adjusted to be closer

together, while words that rarely occur together are pushed further apart.

Mikolov et al. (2013) showed that not only does overall distance in this space

communicate relationships, but that consistent **vector offsets** between words capture important

semantic and syntactic regularities. They wrote:

> "We find that these representations are surprisingly good at capturing syntactic and
>
> semantic regularities in language, and that each relationship is characterized by a
>
> relation-specific vector offset… For example, the male/female relationship is
>
> automatically learned, and with the induced vector representations, 'King − Man +
>
> Woman' results in a vector very close to 'Queen'" (p. 746).

This result significantly expands the utility of word embeddings beyond describing

nearest neighbors, showing that the learned representation encodes relational information that

can support prediction and limited forms of reasoning, such as analogical tasks. In this sense, the

embedding space can be understood as a compressed, high-dimensional summary of word



co-occurrence patterns—analogous to a large word–word association matrix that has been approximated in a much lower-dimensional form, preserving the most informative relationships (Levy & Goldberg, 2014).

While embeddings represent only a single step in the much larger transformer processing, it forms the basis for a number of other applied methods. For example, Abdurahman et al. (2024a) used embedding models to predict human judges' ratings of personality assessment items while Jung et al. (2025) use embeddings as the basis of measuring score reliability in large scale multi-lingual assessment.

### Attention & the Transformer

The Word2Vec embedding is a remarkable map of language, but it is a fixed map. Once training is complete, each word is associated with a single vector, and the relationships between words are fixed in that learned representation. However, the token "bank" means different things in different contexts. In one context, "bank" may refer to a financial institution, while in another it may refer to the side of a river. What language modelers long sought was a way to represent word meaning dynamically, so that a word's representation could shift depending on its context.

This capability emerged with the transformer model originally described by Vaswani et al. (2017) in the influential paper *"Attention Is All You Need."* In transformer models, each token begins with a fixed embedding, and through multiple rounds of the "attention" process, each word repeatedly has its weights adjusted in light of the surrounding words, thereby producing a context-sensitive representation that is then used for next-word prediction. In effect, the model combines broad, general patterns of word usage learned during training with the specific linguistic context of a given sentence at inference time. This supports a balance of general knowledge with situational meaning. During the system development process, the base model



parameters are set by repeatedly feeding the system batches of curated text and running it through the transformer algorithms.  For very large models, this process can take weeks or months, limiting the field of potential developing organizations to a few highly resourced groups.

After a model is trained, new data can be provided in the context as a prompt and new sentences will be generated.  To accomplish this, many transformer-based language models use a complex set of parallel and iterative computations to predict a single next word conditional on what is already in the context. That next word is then added to the context, and the process is repeated. This process of providing text and predicting successive next words using a pre-trained model is called inference, since the system is using learned statistical patterns to infer new outputs rather than updating its internal parameters.

Conceptually, this is analogous to fitting a regression model during training and then using that fitted model to generate predictions for new input data.  However, while most educational researchers will work with statistical models with parameters in the double or triple digits, many LLMs have anywhere from a few million to around half a trillion parameters.  The large open source systems report over a trillion (e.g., Ren et al, 2023), but this is typically in the context of mixture-of-experts architectures which essentially consist of sub-systems that tasks are routed to so that all parameters are not active for any given problem (Zhou, W., et al., 2024 ).

The final output after each iteration is a vector the size of the LLM vocabulary that gives the probability that each word is the most likely next word given the context.  For most words in the vocabulary, this will be essentially zero, but some words will have non-zero values that put them in a candidate set from which the system will choose or sample.  The process for selecting the next word from the candidate output set is typically set with a default approach, though in



some systems, users may be able to select the sampling process by manipulating the 'temperature' setting or the number of tokens considered in the candidate set.

Figure 9 is a screenshot of the on-line interactive Transformer Explainer available at https://poloclub.github.io/transformer-explainer/ (Polo Club of Data Science, n.d.).   It illustrates the flow of information through a transformer model from the submitted text (far left), through tokenization and embedding, attention calculation and the predictive neural networks or multi-Layer Perceptrons (MLPs).  The reader may note this visualization allows for exploration of the details of all 12 attention layers (or heads) and 11 transformer "blocks", each of which is illustrated in the image.  Interested readers are encouraged to experiment by putting different statements (up to 12 words) in the box at the top.



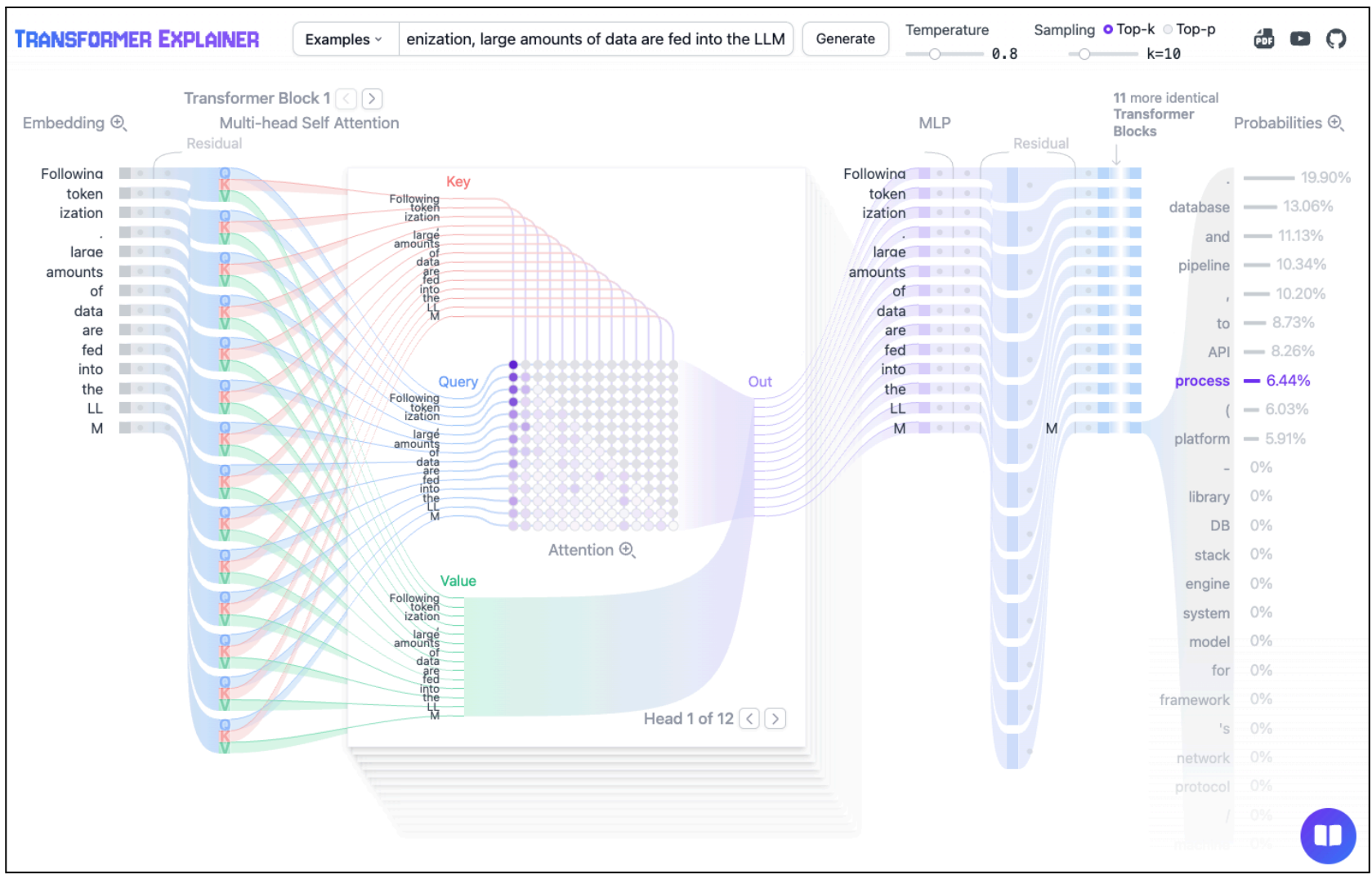

Figure 9. Screenshot of data flow through the Transformer Explainer visualization showing sentence input on the left and next token probabilities on the right illustrating the shape of the probability distribution with Temperature at .8, and Top-k with k=10.



### *Output Probability Hyperparameters*

End users typically cannot affect the overall process described above, but in some interfaces (and APIs) they can affect how the final token is chosen from the set of possible tokens.  Consider the "Probabilities" column on the far right side of Figure 9.  This is the vector of probabilities for each vocabulary token to be the next "word" following the sequence: "Following tokenization, large amounts of data are fed into the LLM".  Here the highest probability token is the period, closely followed by "database","and", and "pipeline".  In this instance, the word "process" was chosen and would be added as the next word in the sentence even though it is a lower probability than a number of others. That is because each word has a probability of being chosen and less likely outcomes also have some probability of being picked, just with lower probabilities.  In this example, we have set sampling to Top-k=10 meaning we only want to consider the top k=10 highest probability words and sum the 100% probability across those.  The reader can see in Figure 10, when we change Top-k to 20, the probabilities spread out across the top 20 most likely tokens and the system sampling from that range to give us the token "API".



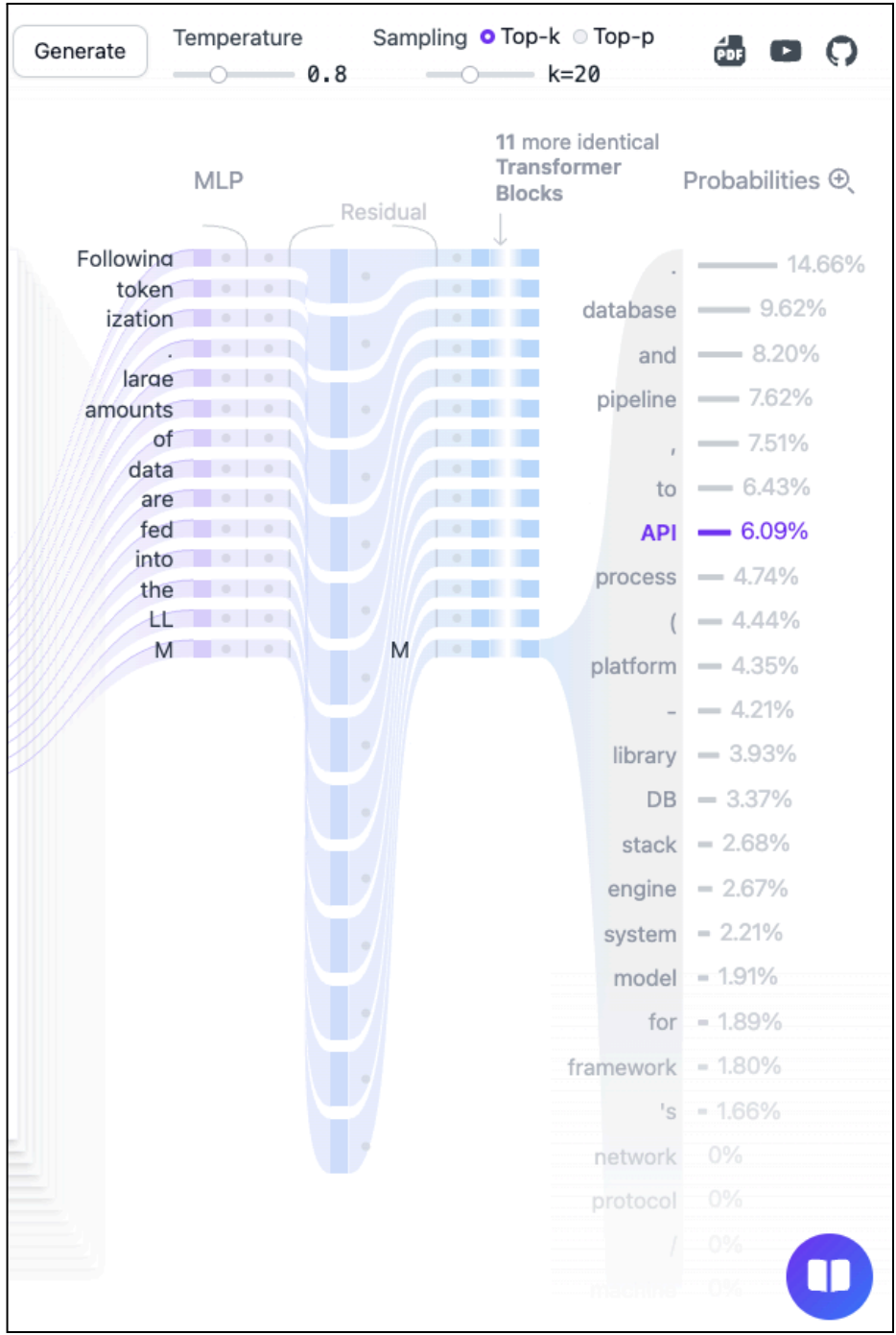

Figure 10. Sampling distribution from transformer prediction with k=20.

Using Top-k we restrict the range of tokens that are considered.  The parameter called

Top-p performs a similar function by restricting the candidates to the top p cumulative percent of

candidates.  For example if one sets p to 50, then only tokens that cumulatively sum to .5 will be



considered. It is similar in impact to Top-k but works in the probability values rather than rank order positions.

Perhaps the most common modification is the use of the temperature setting. Temperature sets the shape of the sampling distribution to emphasize higher probability words (higher temperature) or to make the distribution more platykurtic (lower temperature). The right side of the left most panel of Figure 11 presents an output probability distribution when the temperature is extremely low. In this case only a few probabilities are non-zero, not because the top-k has been set low, but because the distribution has been highly weighted for high probability words, creating a significant skew in the word probability distribution and causing many values to go near zero. The right panel of Figure 11 shows a distribution in which the temperature has been set to a high value and the initial token probabilities are weighted to create a more uniform distribution that allows a wider range of tokens to be more likely to be chosen in the final output. As the reader can see, in this second case the distribution is nearly uniform.



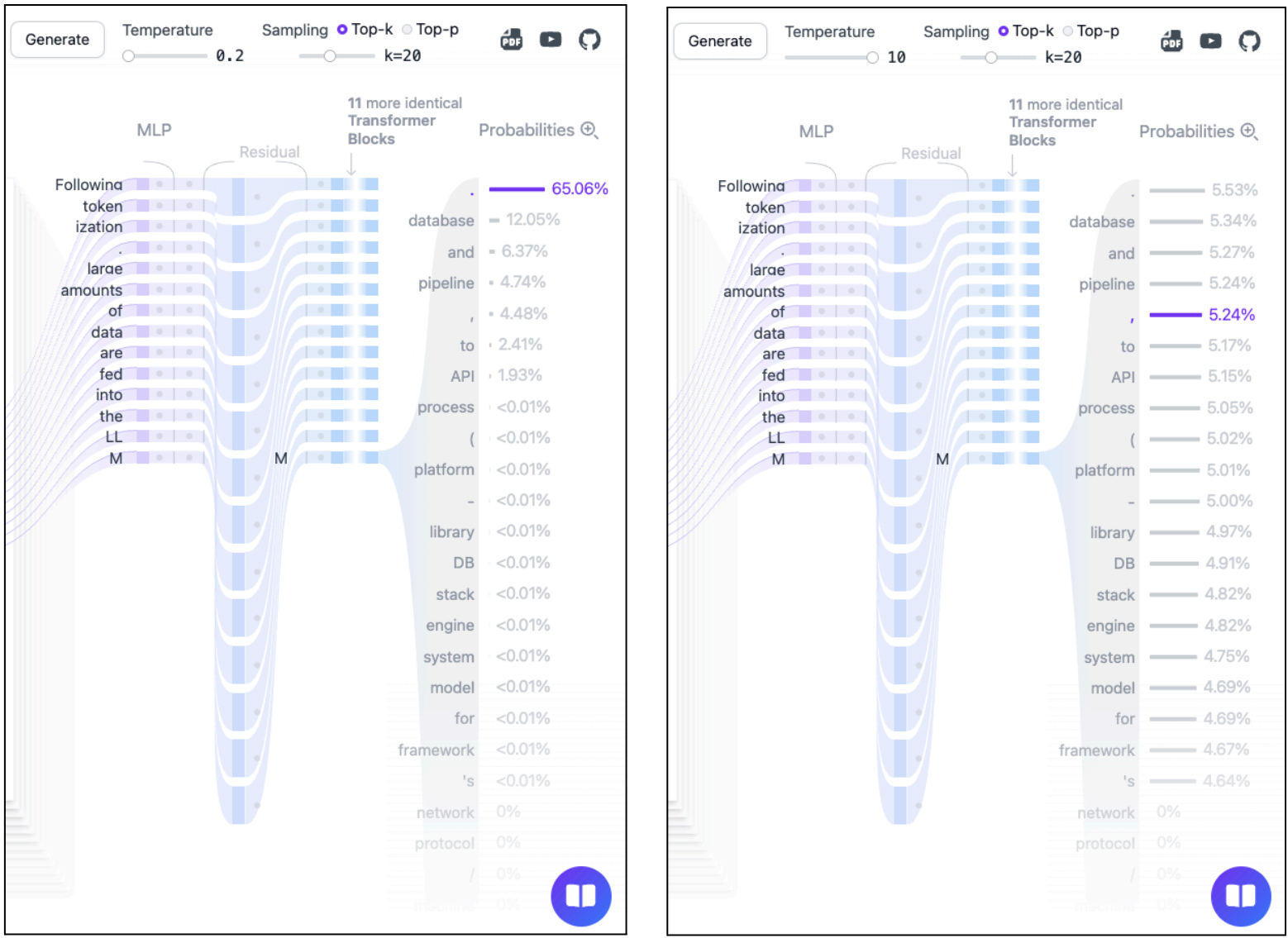

Figure 11. Panels with low (left) and high(right) temperature settings.



Behaviorally, higher temperatures lead to less predictable outcomes and are often applied when creativity is the paramount need while lower temperatures are more predictable and typically used when replicability is the primary value.  However, it is important to remember that unless Top-k is set to 1, there will be sampling from the prediction distribution, and even then there is a chance that results may vary as computing systems typically introduce some random variation as a function of rounding error and differences in training data batch sizes (He & Others, 2025).  Of course, the behavior of the systems needs to be evaluated to ensure both reliability of the predictions and validity of the use.  In production of repeated activities (e.g., item generation), managing temperature or Top-k is often important, though for informal use, it is typically ignored and the system default value accepted.  However, as it is a key determinant of the system behavior, it is an active area of research.  See, for example, Li (2025) for a more detailed discussion with recommendations for different temperatures for different tasks.

Two other notes about temperature.  First, the range of temperatures provided by a system is a product design decision and may vary from system to system though the range 0 to 2 is most common.  Second, in the final steps before configuring the predicted probabilities of each token, a process is required to change the real-valued numbers that come from the neural networks to probabilities.  These mathematical transformations are called logits in the computer science literature.  This is an unfortunate choice as they have no resemblance to logits as traditionally defined in the statistical or psychometric literature and may lead to some confusion when reading the LLM literature.

### Context Windows

Another piece of the architecture that needs to be addressed is the context window.  The context window is the size of the inputs that a system is designed to run through the steps of the



transformer and how many tokens can run through the attention mechanism simultaneously, thereby influencing each other during the predictive inference step.  In this nomenclature, the context refers to inputs given to the system from the user and the context window is the number of tokens that can be processed at any one time.   Sometimes "context" or "context window" are used interchangeably.

The size of the context window is an important determinant of the behavior.  When the user input surpasses the size of the context window, the LLM is not 'attending to' the new information in the same actively updated space as the text that has been pushed out of the context window by the new text, and the behavior of the system may seem irrational.  This was commonplace in the early versions of ChatGPT where some journalists would interact with the system for hours, and the system had no "memory" of the earlier parts of the conversation the way the humans did.  This led to the perception that the system's behavior was inappropriate or even psychotic.  Prior to transformers, "context" was an important NLP concept and was defined as the range of tokens or words on which subsequent predictions were conditioned.

### *Additional Resources*

Given the complexity of the algorithms described above, the interested reader is referred to the excellent animation series provided by Grant Sanderson in the 3Blue1Brown YouTube channel.  Sanderson expands a series of videos explaining the logic of  Deep Learning with videos on the Transformer Architecture (Chapter 5; Sanderson, 2024a), Attention in Transformers (Chapter 6, 2024b),  and How might LLMs store facts, Chapter 7, (Sanderson, 2024c).  Sanderson (2024d) provides a striking 8 minute video that gets to the heart of the points discussed above.  For those interested tutorials on the computational steps of the transformer computations, the readers are referred to the excellent interactive treatment by Yeh (2025)



**Model Post-Training**

After a model has gone through the pre-training regime described above, it is typically considered a "base model" and "generally ready" but not necessarily ready for a specific range of tasks. As Ouyang et al. (2022) put it "predicting the next token on a webpage from the internet—is different from the objective "follow the user's instructions helpfully and safely"" (p. 2), the common starting instruction for LLMs at the time. This is analogous to a young child who has a form of general intelligence but does not have the specific knowledge required for some tasks.

To close this gap, developers typically add two phases. The first is to give the system specific examples via supervised fine tuning (SFT), sometimes simply referred to as Fine Tuning (FT), or Instruction Fine Tuning when focusing on learning to follow human instructions (Zhang, S., 2024; Yang et al., 2025). The second (and subsequent) approach is to provide reward signals based on human ranking of system behavior to reinforce certain behavioral leanings by applying Reinforcement Learning with Human Feedback (RLHF) or sometimes simple Reinforcement Learning (RL). These are rich topics and active areas of research, especially RL so the treatment in this section will be necessarily lean..

***Supervised Fine Tuning***

As noted above, SFT, also called instruction fine tuning (Zhang et al., 2024) is the process of providing the models with specific examples to help strengthen the associations in a domain in order to improve or focus model behavior. It is important to note that "fine tuning" in this context is a computational process in which the weights of the network are updated. This creates a change to the inference architecture that is built into future inference and not the same as giving transitory information as occurs when examples are given within prompts. Given this



specific and commonplace definition of 'fine tuning', the reader should avoid using the term for other general processes such as improving system performance through prompting in which case "iterative improvement" or other expressions should be used.

Ouyang et al. (2022) were the first to demonstrate the FT process during the creation of the Instruct-GPT system. Given only its base GPT-3 model from pre-training, it was prone to hallucinations, toxicity and failure to follow user instructions. To address these concerns, Ouyang et al. provided approximately 13 thousand example question / answer pairs aligned with the desired behavior of the system to serve as "instruction following examples" (OpenAI, Jan 2022). This updated model was considered the "instruction model" which was then refined with examples of human dialogs via RLHF to create the final Chat-GPT (OpenAI, November 2022).

In addition to improving the specific behavior of large models (such as GPT-3), the most common use of SFT is to customize open models to build specific behavior such as scoring of computer programming (Ross et al., 2025) or scoring language performances (cf. Foltz et al., this volume). Prior to 2025, the only option for researchers to conduct SFT was to use an open model with access to the weights that could be changed. Since then, OpenAI (OpenAI, n.d.) has developed tools to support end users to conduct FT on their models. Zhang, S. et al. (2024) provide a more detailed overview of the various systems and approaches to FT in this context.

For most researchers, however, customized fine tuning of models becomes less relevant as foundational models improve in their capabilities as implemented through prompting. In the early years, the limitations of LLMs required researchers to customize models to enhance behavior in specific domains (a sort of 'sending the model to school for a specific topic') though this need will continue to drop over time. See, for instance Foltz et al. (this volume) discussion



of transitioning from fine tuned to prompt-only approaches and the report from DiCerbo et al. (this volume) that their global deployment at Khan Academy is entirely prompt based.

*Reinforcement Learning*

Reinforcement learning is a process of updating model weights on the basis of system behavior on tasks. Initially this largely occurred as Reinforcement Learning with Human Feedback, in which human labelers would rank answers provided by LLMs according to appropriateness or correctness (e.g. Ouyang et al., 2022). This has been an area of very active research and evolved so that systems are receiving judgement signals both from humans as well as from the information the system may receive when attempting other activities. That is to say, another LLM can act as a judge on the correctness signals from pre-fabricated benchmark tests that can be used to help shape an LLM's downstream behavior. This changes the description of what the LLM does from "predict next words based on the statistical patterns found in the internet and related digital data" to something more like "Given the statistical distribution of words in the vast public corpus of human language, what words that users and raters would most approve of are most likely to follow the sequence?" (Shanahan, 2024, p. 68).

Reinforcement Learning is not a language-model specific AI technique but a general purpose technique being widely explored in the area of language, image interpretation and robotics. Both RL and FT are often seen as processes to help LLMs "align" more specifically with the goals of the user than the general goal of overall next token prediction. However, there are often unintended side effects in which the more superior performance on specific tasks that occur because of the specificity of FT and RL may occur at the cost of performance in some aspects of the base model – a condition called the "alignment tax" (e.g. Lin et al. 2024). Wu et al. (2024) have argued that subgroup and behavioral misrepresentation may be another



deleterious side-effect of RLHF (Wu et al., 2025). The main point to take away here is that while recent advances continue to lead to dramatic reductions in hallucinations and improvements in many specific task areas, all these techniques are very new and changes in understanding their statistical dynamics and applied impacts are ongoing. The interested reader will find more information about impacts of RL in Zhang, Y., et al. (2024) and consult Maura-Rivero et al. (2025) for examples of emerging alternatives.

**Reasoning Models**

A special case of SFT/RL is the use of such training to create "Reasoning Models". Reasoning models are models who have been post-trained with systematic examples of how to think through problems step by step, typically by giving examples of chain of thought thinking (Wei, et al., 2022). For example, Jha et al. (2025) provided over 300,000 problems from various domains whose answers were annotated in a step-by-step manner, leading to high performance in math and reasoning tasks of dramatically larger models.

Using these and similar techniques, almost all open and proprietary organizations have incorporated reasoning into their system architectures including the latest Gemini and GPT systems. Reasoning models have been shown to be particularly effective in addressing tasks that are 1) highly ambiguous, 2) include large amounts of unstructured data, 3) require finding nuanced relationships across large datasets, 4) multistep agentic planning, and 5) visual reasoning on images, diagrams etc. (OpenAI, n.d.b). Sprague et al., (2025) conducted a meta-analysis of 100 CoT papers and ran evaluations on over 20 datasets and 14 models. They concluded "the primary benefit of CoT comes in the ability to execute symbolic steps and track their output. Not all tasks have this feature: for example, questions from CommonsenseQA can hardly be translated into formally grounded and executable solution plans" (p. 10).



Reasoning models have evolved quickly over the last few years, with strategies such as self-consistency checking (Wang et al., 2023) in which multiple thought chains are developed and a modal response is taken as the appropriate answer. This continues to be an active area of research with rapid advances in the last 2 years and implementation in many large scale systems such as ChatGPT 5 and Gemini 3.0.

**Synthetic Data & Model Training**

As data for training models becomes increasingly important, researchers have turned to LLMs themselves to generate data for subsequent training. For example, Zelikman et al. (2022) extended this idea into a sustainable and scalable generative system in the context of CoT tuning. The approach is to give the system a small set of initial CoT examples to teach it the approach, and then have it explain the answers to many questions. Next, identify the correctly explained rationales and fine tune that high-quality rational base repeating the process. The model can then generate the next set of training data iteratively. As the authors note, "This is a synergistic process, where improvements in rationale generation improve the training data, and improvements in training data further improve rationale generation" (p. 2).

This is just one illustration of the larger trend of using  is often called system self-instruction (Roy, 2025; Wang, Y.,  2023) and was the approach behind the launch of the DeepSeek-R1 (Deepseek-AI et al., 2025) in January of 2025 that set off a significant financial market reaction when it was realized that synthetic data (created by the LLMs themselves) together with RL may lead to significantly less expensive models than previously envisioned.

**Guardrails and System Prompts**

Having constructed the basic underlying determinants of the system behavior, additional steps are taken to constrain the system behavior: Guardrails and system prompts. Guardrails



refer to policies and their implementations, both technical and organizational, that aim to prevent or mitigate undesired system behavior (NIST, 2023).   There are a number of ways to accomplish this, including the use of system prompts, which are addressed here first.  Keep in mind, however, that to appropriately  implement guardrails comprehensively, organizations need to consider appropriate use and safety throughout the product lifecycle – from goal specification through all the steps described in preceding sections.  The best protection is a system which was developed to closely match the desired behavioral profile and for which-post hoc guardrails and error mitigations are not required.

The system prompt (aka system message), is a prompt, often of considerable size, that is loaded at the time the user starts an interaction with the system.  It is essentially a pre-established prompt that constitutes the start up instructions for that user session.  Given this important role, they define "the general instructions, safety guidelines, and constraints for the LLM, as well as tools available to" the GAIS (Wallace et al., 2024). In August 2024, Anthropic became the first major vendor to publish system prompts with versioned release notes. Anthropic's prompts explicitly balance warmth with honesty: "Claude uses a warm tone. Claude treats users with kindness and avoids making negative or condescending assumptions… Claude is still willing to push back on users and be honest, but does so constructively—with kindness, empathy, and the user's best interests in mind" (Anthropic, 2024). The value of such proactive specification became evident in April 2025, when OpenAI rolled back a GPT-4o update after users reported excessively fawning responses; the fix required adding "avoid ungrounded or sycophantic flattery" to the system prompt. Language addressing behavior that OpenAI's own Model Spec already discouraged but that training on short-term user feedback had reinforced (Leswing, 2025; Ng, 2025; OpenAI, 2025; Willison, 2025).



Microsoft publishes safety system message templates addressing harmful content and grounding instructions (Microsoft, 2024), while xAI has released Grok's prompts on GitHub (a software sharing site), though independent researchers discovered in July 2025 that Grok 4 frequently searches Elon Musk's posts before answering controversial questions, with its chain-of-thought reasoning explicitly stating it was "searching for Elon Musk views" on topics like immigration and the Israeli-Palestinian conflict (Howard, 2025; Zeff, 2025).

Typically system prompts are standardized across all users, but some systems also allow users to specify their own additional system message as well. For example, users of the free gmail platform have access to a range of generative AI systems including the Google AI Studio at aistudio.google.com. That interface allows for numerous customizations including adjusting temperature and top-up as well as adding a personalized system message to supplement the system-wide message. Importantly, the system prompt may be loaded by the end-user interface for chat-based interactions and do not impact the behavior of the system when accessed through the API (e.g. Anthropic, 2024). The reader may find additional artifacts in the website https://github.com/jujumilk3/leaked-system-prompts (gyudoza, 2025), which purportedly publishes proprietary system messages. The reader should consider the provenance of the leak and potential acknowledgement of the parent organization when assessing the veracity of the supposed leak as forgeries cannot be otherwise ruled out.

Aside from using system prompts, to the degree possible, developers will often build in guardrails in the post-training phase to embed them in the product. For example, Bai et al. (2022) describe a process of using RL to train systems to identify and deflect harmful queries in the Anthropic Claude systems. A common pattern here is to give examples of inappropriate



queries with the corresponding appropriate response from the LLM. A second method is to work to steer control of the system behavior via the system prompt, also called the system message.

In addition to the preventative measures embedded in post-training and system prompts, ongoing evaluation and mitigation is also required during system use (aka 'inference'). In this third approach, systems are designed to filter and flag inappropriate output after it has been drafted (and before it is shared with the user). This could rely on explicit rules or rule patterns, or use machine learning techniques to predict whether a behavioral pattern is aligned with clusters of inappropriate behavior. There is a thriving industry and active research on the development of software to undertake this final step (e.g., Han et al., 2025).

The guardrail infrastructure is the last chance to catch unwanted system behavior and redirect or eliminate it. To get a sense of the range of issues that must be addressed, we can look to the "red teaming" literature that documents the range of mischief dedicated (red) teams can create when testing LLM systems. Ganguli et al. (2022), for example, clustered red team attack points into clusters that included successfully soliciting personally identifiable information, doxing, cheating, misinformation, adult content, making and selling drugs, soliciting advice on violence, and harmful health information. Additional concerns that must be addressed include biases of many types, and a range of security issues, alignment with organizational economic goals (e.g. don't recommend the competition), legal and regulatory requirements (e.g. FERPA), safety and harm prevention, cultural appropriateness and language requirements. Of course the goal is to identify potential issues so that systems can be updated to address them. Systems developers keep in mind that such breaches can happen based on either user ill intent or system mis-alignment.



## Tensions

### *Hallucinations & Expectations*

So far, we have covered a wide range of practices that impact GAIS behavior including data collection and curation, model creation via tokenization, embeddings, attention and prediction, FT, RL, system messages and other guardrails. This covers the core of the original LLM dynamics from earlier large models though numerous extensions exist that will be discussed in the next section.

Here, let us pause to consider that these computational mechanisms lead to the fundamental system characterization with which we started the chapter: A probabilistic word predictor that generates text based on the statistical patterns found in training data and the prompt (aka "context") that the user provides. As Bender et al. (2021) put it, the systems are "Stochastic Parrots" or as Hicks et al. (2024) titled their essay, "ChatGPT is Bullshit", arguing that "because they are designed to produce text that looks truth-apt without any actual concern for the truth, it seems appropriate to call their outputs bullshit" (p. 37). See Gunkel & Coghlan (2025) for a more detailed analysis of "The Bullshit Literature". This is all exacerbated by the visceral experience of interacting with these systems where, as Shanahan (2024) notes, "Interacting with a contemporary LLM-based conversational agent can create an illusion of being in the presence of a thinking creature. Yet, in their very nature, such systems are fundamentally not like us." (p. 68).

As we have seen, contrary to common belief, these are not traditional information retrieval systems that "look up" facts in databases as would be the primary deterministic approach. Rather, they are probabilistic predictors only as good as their data and training, and hence infallible, and in some contexts, unpredictable. When these systems fail to meet our



expectations, we often say they are "hallucinating" (Huang, et al, 2025). Hallucination is most frequently discussed in the context of failure of information retrieval, that is "to give the right answer". We can imagine a number of reasons for this to occur including absence of the information in the training data (Kalai & Vempala, 2024), ineffective model training (Sun et al., 2025), improper behavior because of fine tuning effects (Kalai et al. 2025), or competing information in the data itself (Zhang et al, 2024). There is active research in the computer science community aimed at all of these concerns as well as work to add additional information to the systems by giving the systems access to tools such as archival data (RAG: Gao et al., 2024) or the WWW. Hong et al. (2024) proposed an online "Hallucinations Leaderboard" that can be accessed at https://huggingface.co/spaces/Akash687/leaderboard as of December 2025.

We must remember too that system behavior is also driven by the content of user prompts, requiring us to consider "user error" as a significant source of the combined human-computer system failure as well. Finally, it is important to remember that these systems are not only used for information retrieval, but are used to generate a wide range of texts including poems, fiction, hypothetical scenarios and personalities, and many other tasks for which there are no "correct" answers and in fact the human goal of creative hallucination is actually desired.

### Impacts of Synthetic Data

Another area of concern is the repeated use of synthetic data from the "self-instruct" process. Here, several authors have raised concerns that over time the iterative re-training of systems on their own data will lead to degradation of quality to the point of "model collapse" (Shumailov et al., 2024, Wenger, 2024), or in the context of image generation, models go "MAD: via "Model Autophagy Disorder" (Alemohammad et al., 2023). Schaeffer et al. (2025) argued



for the importance of a nuanced discussion about the problem and suggested a taxonomy of what "model collapse" means to authors in different papers. While the paper is an excellent introduction to the issue and is recommended for additional reading, two points are important to highlight here. First, Schaeffer et al. (2025) cite OpenAI CEO, Sam Altman, as reporting that as of February 2024, "ChatGPT alone produces 1/1000 of all words produced by humanity each day" (Schaeffer et al., p. 6). The authors note both that this is a significant amount of data (and does not account for changes since then or the rise of additional systems such as Google Gemini) and that traditional sources of data continue to grow as well, potentially keeping some balance between types of sources.

Second, referring to infrequent data that would be observed in the "tail" of a histogram or probability distribution, Schaeffer et al. note that "**Real tail data and modes will be lost, but how many and how quickly is unclear**. Loss of diversity is a real issue, with disproportionate harms often born by subgroups."(p. 9). [Emphasis in the original]. The idea is that when data are generated from GAISs themselves, low data (in the tails) linguistic forms and communities will iteratively become lower and lower. This should be flagged as a significant potential social ill that could exacerbate the bias, stereotyping and identity flattening that already exists in current implementations (Blodgett et al., 2020; Lee, et al. (2024); Shelby et al., 2023) and agree with Schaeffer et al. that additional research needs to be undertaken along these lines whether caused by model collapse or model construction. See Zhang et al. (2025) for an example of research in this emerging field.

**LLMs Are Not LLMs Anymore**

The discussion to this point has focused largely on the domain of text-to-text (that is text inputs and text outputs) models that undergirded the original LLMs. This was undertaken in



order to clearly communicate many of the foundational concepts in GAISs without overwhelming the reader with complexity.  However, it essential for the reader to know that as of the writing of this paper, (December, 2025) the most prominent models (e.g. GPT-5 and Gemini) are in fact not LLMs, but rather Large Multimodal Models (LMMs) or sometimes referred to as Multimodal Large Language Models (MLLMs, Caffagni et al., 2024,  Yin, et al., 2024).  These systems extend the logics of LLMs explained above by training not simply with text, but also variously with images, text, and possibly sound or video.  This allows for striking interactions between these representations moving from text-to-text in the LLM to X-to-X in the LMM where X equals text, images, video, and sound.  For example, Google's Gemini platform is quite adept at performing search within a video.  Simply upload a video file and ask a question about a scene in the video such as "where in the video does X appear " where X is a character, scene feature, text overlay or dialog.  This provides free access to video search technologies that would have been available to very few users just a few years ago.

Similarly, both ChatGPT and Gemini have become proficient at recognizing and interpreting images and formula in ways that allow them to work with highly visual representations common in mathematics and scientific curricula and practice (cf. Liu et al., this volume) and overcome prior limitations in transcribing handwriting, interpreting equations, writing code to address problems presented as images,  and even some aspects of visual reasoning (Yin et al., 2024).   While these systems have generally been available since mid-2024, the overall quality of these systems provided both by OpenAI's GPT series and Google's Gemini series have shown dramatic improvements in 2025.  Küchemann et al. (2025) describe a range of opportunities, challenges and mitigations for the impact the widespread availability of such systems including leveraging the multimedia effect discussed by, for example, Mayer (2005);



enhancing collaboration (e.g. Foltz et al., this volume), enhancing explainable and accessible user interfaces, and bridging explainability gaps between visual, audio & textual representations. Küchemann et al. also point out directions in which educational research may be enhanced including in conducting literature reviews, creating multimodal assessments and learning materials, supporting advanced coding and data interpretation, and opening the possibilities for a range of text-to-speech and other x-to-x based learning tools.



Figure 12. Screen shots of the ChatGPT mobile app taken spring 2025.



Consider Figure 12, which is a screen shot of the ChatGPT mobile app. using the camera to take a picture of a diagram on a whiteboard with the simple prompt of "Answer that question" and appropriate computation and explanation in response. Image taken spring 2025. This process is enabled by the multimodal nature of the LMM which has tokenized and applied attention to textual and image samples simultaneously.

Because the most interactive forms of LMM deployment are on mobile devices, we have seen dramatically higher use of LMM features among students than among faculty, with many faculty unaware that the fundamental range of functionality of generative AI has changed over the last two years. This will likely exacerbate the experience and perception gaps that already exist between students and teachers and needs to be considered carefully in systemic roll out of AI-related professional development. Further, these technologies are extremely nascent and use cases for them are expected to grow exponentially in the coming years.

**The Privileged Properties of Computer Code**

In light of all we have discussed up to this point, one general pattern emerges: LLMs and their variants will perform better when they both have a lot of data relevant to a particular task, and the completion of the task is highly predictable. This is especially true in earlier models that relied primarily on next-token prediction. One area where this is especially effective is in support systems for computer coding. Computer code is digitally native and there are many shared repositories for computer code (e.g. Github; github.com) that individuals use to share their work as electronic portfolios. Likewise, all open source repositories can serve as training data from LLMs. When these large repositories of training data are mixed with the task structure



of computer coding that is highly predictable (at least relative to human language), it creates a unique combination of circumstances that are well aligned with the strengths of LLMs.

Accordingly computer code generation has been one of the most prominent success stories in LLM applications and will likely continue to be so going forward.  This has important implications for the extensibility of GAISs.  In the same way we saw potential for data generated from LLMs to train subsequent LLM generations, computer code is simply symbolic sequential data in the same way that human language is..  Accordingly, there are current workflow patterns in which examples of computer code or other structured documents, can both be fed into LLMs to teach them how to code, as well as generate new examples for subsequent retraining.

These features of the GAISs and the internet enabled data ecosystems are leading to cases where human involvement in complex coding is slowly receding, thereby upsetting longstanding traditions for teaching and workforce development in computer science.

## LLMs Are Not (Just) Models Anymore

While LLMs are no longer LLMs, but rather LMMs, LLMs are almost universally not simply 'models', but rather product ecosystems.   For example, ChatGPT users today are faced with a complex interface that reflects the complex set of features offered.  These include features such as file upload, web search, python code writing and execution, image generation, command bundling in GPTs, projects, access to different model running modes, and so on.  Tables 3 and 4 provide a summary overview of many of the common features available in December 2025 along with descriptions of the variation in the three most common commercial application product lines: ChatGPT by OpenAI, Claude by Anthropic, and Gemini by Google.

Rather than provide a comprehensive guide, we hope bring to the forefront the idea that when individuals are working with generative systems, they are dealing with a complex product



that elevates some key technical features (e.g. deep research) but also working with a complex

Software as a Service (Saas) framework that also has additional hidden features such as safety

guardrails, security management features, user management features (e.g. teams), memory

management etc.  For marketing purposes these systems are often referred to in terms of their

foundational model version (e.g. ChatGPT 4.0 vs 5.0) which do differ in their capabilities, yet

users work not only with the model, but in a complex interaction between the model and the

other system features.

     Table 3  provides an overview of features supporting the inference or generative use of

the system.  Perhaps the most important feature in this area of concern is the context window.  As

the reader can see, depending on the model variation within a vendor, all vendors have some

versions with context windows of 1 million tokens.  Likewise, all vendors allow extensive file

upload with corresponding interpretability of text and image with Gemini including video as

well.  The addition of these features is significant.  Earlier models suffered from limited context

windows that would lead to a point at which the systems would lose track of their earlier

behavior and act in ways inconsistent with prior dialogs.  File uploads and prior-chat memory

allow the systems to act with more consistent behavior that creates a sort of "social memory" that

persists between interactions.  Availability of such features likely varies across different product

versions or mode of sale (e.g. free or subscription).



| Feature | ChatGPT | Claude | Gemini |
|---|---|---|---|
| Context Window | 128K-1M | 200K-1M | 1M+ |
| File Upload | 512MB; Code Interpreter processing | 50MB; 500+ page docs | 1,500+ pages plus video Drive integration |
| Memory | Automatic, persistent | On-demand, transparent | Google account-based |
| Web Searchability | Yes | Yes | Always-on (Google) |

Table 3. Features Supporting Enhanced Inference

Web searchability is also a significant add on to the base LLMs as earlier versions were locked into information up to their training date. Now, each of these systems natively searches the internet for information before returning answers. The positive side to this is the potential for supporting web searches per se, as well as the possibility that answers can use very up to date information. The negative side is that now answers are a mix of processes from pre-training, post-training and web searching. When a response is generated by these systems, it is now difficult to judge the thoroughness of and validity of the response because it is not clear if the web search was complete or useful and how the response is mixing the different sources. Is the response an up to date response equivalent to a traditional google search using only web indexing? Is it better than a traditional google search because it combines real-time search with



deeply processed LLM information, or is it worse given that the whole process is more opaque? Standards for evaluation and acceptance will evolve over time, but in the meantime, it is important for users to keep in mind that the mix of the different sources makes it difficult to interpret or trust comprehensively.

| Feature | ChatGPT | Claude | Gemini |
|---|---|---|---|
| Code Execution | Python (server-side) | JavaScript/ Python (Claude Code) | Limited (Colablink) |
| Image Generation | Yes (GPT Image 1.5 ) | None | Yes (Nano Banana Pro ) |
| Voice/Video | Voice Mode; limited video | No | Gemini Live; native video |
| Workflow Config | GPTs + Actions + Store | Projects Claude Skills | Gems (personal) |
| Workspace Management | Canvas (editing focus) | Artifacts / Teams (preview focus) | Google Integration (Drive, Docs, Sheets) |
| Ecosystem Connection | Tools Webbrowser | MCP Agentic Sharing | Google integration |

Table 4. Features for Advanced Output Modes

Table 4 provides the status of a number of features related to advanced output modes for each of the three most prominent proprietary systems as of December of 2025.  We start here



with the code execution feature. This feature was first introduced by OpenAI in GPT-4. In early versions the LLM system would determine if the query from the user was best answered by standard LLM prompting or by running computer code. For example, a common demonstration in 2024 was to ask ChatGPT "what is the square root of 100" to which it would quickly reply with "10". When asked the square root of a large random number, the system would typically open the python interpreter (an environment for running programs written in the Python computer programming language), enter Python code to solve the problem and then return the answer. Figure 13 is a screenshot of GPT-5.2 showing its work to convince the author of the answer.

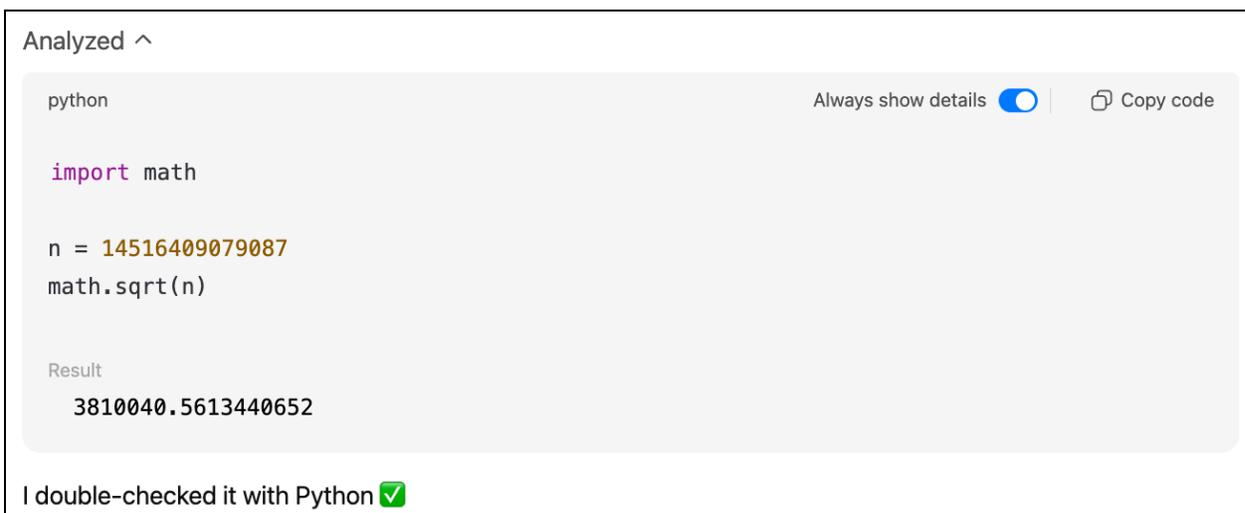

Figure 13. Screenshot of ChatGPT 5.2 running Python code through the user interface to demonstrate the square root answer for a large random number.

Running code is a part of a large construct of GPT-Actions and the idea of giving LLMs access to "tools" writ large. Tools are any software that can act in the world such as opening and reading your email, passing the email with a prompt to summarize it, passing the answer back to



the email system and so on.  What this all means is that in the last few years these systems have rapidly moved from novel chatbots to active software systems that may be allowed to act in complex software environments by retrieving, acting on, and sending data to behave in our computing systems on our behalf.  While on the positive side, this may lead to extreme gains in personal productivity and rapid product development, it also opens a pandora's box for unintended actions and significant privacy and security breaches.

With regard to workspace configuration, ChatGPT series has a feature called "GPTs" and Gemini have a system called "Gems" in which users can enter prompts and background data that can be called up in the future with automatically loaded context and system messaging.  This supports wide ranging re-use and sharing of extended prompts as code snippets.

In terms of workspace management, both ChatGPT and Claude have spaces in their dedicated desktop software that provides real-time rendering of outputs of the system outside of the prompt window called the canvas or artifacts ChatGPT and Claude respectively.

It is important to note that these systems are routinely developing various tool connection systems so the LLMs can interact with other software systems.  Anthropic established an end-system agnostic translator between LLMs and other systems called the Model Context Protocol (MCP) which is rapidly becoming an industry standard.  It allows various LLMs to talk seamlessly to a software environment (say an email client, or database) by providing a translation layer so the MCP can translate between human language used in the prompts and the programming language the end system needs to work in.  This is rapidly becoming an industry standard, greatly accelerating the move toward connected LLMs seamlessly with existing systems.



Both OpenAI and Anthropic are also working on giving LLMs control of a web browser potentially allowing it to behave on the internet on your behalf. While the appeal of a system performing such tasks as finding a book or article in your library sounds appealing, the fact that the system would likely be managing your user ID and password to support authentication, opens a large set of problems for managing security and privacy protection.

**Summary: LLMs as Complex Computational Ecosystems**

In this section we have described the fundamental components and characteristics of LLMs. Originally "only" probabilistic word predictors, they have evolved into complex mulit-modal systems that integrate with other computational systems to model and generate natural and man-made languages with increasing accuracy. Nevertheless, given their non-deterministic and opaque provenance they remain a constant risk for exhibiting undesirable, unintended, or undetected aberrant behavior. With greatness comes responsibility, yet the mechanisms for ensuring the responsibility are lagging behind the (unreliable) behavioral sophistication.

## Prompts and Other Inputs as Determinants of System Behavior

Now we shift our focus from how the systems work, to how the end user should work it. While the product and model features described above determine the range of potential behaviors of an LLM, the data entered into the context window at inference run time, the prompt, determines more precisely the specific behavior that is outputted. The product/model is the boat, and the prompt is a loose rudder: influential, but imprecise.

This section is organized as follows. First, a brief discussion of the interplay between the context window, in-context-learning (ICL) . Second, a discussion of some of the major techniques for prompting organized into the clusters of Identity Specification, Information



Provisioning, Process Specification & System Interaction Management.  This section concludes with a discussion of the Prompt Engineering (PE) workflow and its implications.

**The Unreasonable Effectiveness of Prompts**

While the term 'prompt' is inextricably related to 'LLM' for many GAI users, like many features we have seen so far, the pivotal role of the prompt was, in some ways, unanticipated:

"GPT-3 [Brown et al. 2020], with 175 billion parameters compared to GPT-2's 1.5 billion, permits in-context learning, in which the language model can be adapted to a downstream task simply by providing it with a prompt (a natural language description of the task), an emergent property that was neither specifically trained for nor anticipated to arise" (Bommasani et al., 2021, p. 5).

It is important to note the expression of "in context learning" (ICL) that explicitly calls out the idea that the LLM behavior is "learned" from within the 'context' of inputs after pre- and post-training.  This was considered a watershed insight as prior efforts at evoking LLM behavior were focused almost exclusively on FT, a process that required programming and data science expertise.  Liu et al. (2021) wrote about a "sea change" in method noting: "The advantage of this method is that, given a suite of appropriate prompts, a single LM trained in an entirely unsupervised fashion can be used to solve a great number of tasks" (Liu et al., 2021, p. 2.).  The authors were prescient in anticipating a new subdiscipline:  "However, as with most conceptually enticing prospects, there is a catch–this method introduces the necessity for *prompt engineering*, finding the most appropriate prompt to allow a LM to solve the task at hand."(Liu et al., 2021, p. 2.).



**Approaching Prompting**

Despite the term 'prompt engineering' being less than 5 years old, various interpretations and confusions have evolved and no single viewpoint holds universally. For the purposes of this paper we use the following definitions:

- Prompt - Data entered into the context window (during inference) to elicit system behavior

- Prompting - The process of creating and entering prompts

- Prompt Engineering - The systematic analysis of prompt requirements, prompting implementation and evaluation with appropriate iteration.

This nomenclature considers all activity related to entering prompts into a system as prompting, while reserving the label "prompt engineering" for more formal, evaluative and iterative processes. It aligns with other common definitions (e.g. Schulhoff et al., 2025).

It is valuable to approach prompting with several core beliefs built on the information discussed above. These core beliefs should include:

1. The system is a probabilistic string predictor and the goal of prompting is to elicit specific token-prediction behavior in the system by providing prompts aligned with the users goals and system characteristics.

2. The system does not think or understand as humans do. In prompting we are simply trying to elicit linguistic behavior from a model.

3. Every interaction is concerned with building context to evoke LLM behavior. Specificity and linguistic association are the key tools for prompting.



4. Typically the system has been tuned for iterative interaction across the context window.  Accordingly, iteration is often essential.

5. The behavior depends on the system, the prompt and the task.  The system is nearly constantly changing and so all prompting remains an empirical endeavor that requires an appropriate evaluation framework.

The fight with anthropomorphism is difficult.  On the one hand, as the systems have been trained to imitate humans insofar as surface level language is concerned, it is helpful to "be yourself" and interact with the system with iterative feedback and detail the way one would with a colleague or intern.  On the other hand, one must remember the system does not "understand" intent and context the way a human does and that specificity is essential (Shanahan, 2023).

**Prompting Strategies and Techniques**

After ICL was discovered as a viable alternative to simply fine tuning all models, a race ensued to understand which prompting forms were most effective and long lists of techniques are available. Here, this section is organized around several categories that serve as organizing strategies and discuss prototypical techniques within those categories: 1) Identity Specification, 2) Information Provisioning, 3) Process Specification, and 4) System Interaction Management. This structure is an approximation of the terms used by Behrens (2025a) in course material (Orchard et al., 2025) which provides an extensive taxonomy and examples beyond what is possible here.

***Identity Specification (Persona Prompting)***

*Identity specification* refers to prompt designs that instruct a language model to adopt a particular role, identity, or perspective (e.g., "act as a historian" or "you are a medical expert"). This technique is often described as *persona prompting* or *role-playing* and was framed by



Shanahan et al. (2023) as a core form of role-based interaction with LLMs. Identity specification can meaningfully affect model outputs, but its effects are highly task-, model-, and alignment-dependent.

**1. Align persona specification tightly with task goals.**

Persona prompts are most effective when the specified identity directly supports the task being performed (e.g., *"You are an expert in Minoan civilization"* for a question about Bronze Age Crete). Across studies, narrowly tailored expertise attributions produce positive or neutral effects, while broad or generic personas (e.g., *"expert in history"*) often show no benefit (Luz de Araujo, 2025; Zheng et al., 2024). Task<–> persona alignment is therefore a primary determinant of effectiveness.

**2. Prefer task-relevant expertise over demographic or social identities.**

Specifying levels of expertise or domain knowledge generally yields more reliable improvements than assigning demographic or social identities (e.g., race, gender, religion, disability). Studies of socio-demographic prompting show that such identities can influence outputs on subjective judgment tasks (e.g., toxicity classification), sometimes improving alignment with human annotators, but effects vary substantially by model and prompt formulation (Beck et al., 2024). In contrast, demographic personas have also been shown to introduce bias and performance degradation on many tasks, particularly in earlier models (Gupta et al., 2024; Salewski et al., 2023).

**3. Avoid extraneous or unrelated identity attributes.**

Adding identity details unrelated to the task (e.g., *"your name is…"*, *"your favorite color is…"*) may degrade performance if the attribute is unrelated to the expected behavior. . Large-scale evaluations show that such extraneous attributes introduce noise and can negatively



affect accuracy, with especially strong negative effects for personas associated with lower educational attainment or unrelated personal characteristics (Luz de Araujo, 2025). Identity specification should therefore be minimal and purpose-driven.

**4. Expect stronger effects on subjective and creative tasks than on formal reasoning.**

Persona prompts have larger and more consistent effects on tasks involving subjective judgment, interpretation, or creative expression than on tasks such as mathematical computation or formal logical reasoning. Empirical studies show little to no benefit of persona specification for reasoning-heavy tasks, while creative or interpretive tasks often benefit substantially when personas are well aligned (Wang et al., 2024).

**5. Account for model sensitivity and generation strategy.**

Persona effects vary across model families and generations. Earlier models are more sensitive to identity specification and more prone to stereotyping or performance collapse under poorly chosen personas, while newer models tend to be more robust but still benefit from carefully designed, task-specific personas (Gupta et al., 2024; Luz de Araujo, 2025). Systems that dynamically generate or select personas tailored to the task domain show consistent gains over static persona prompts (Wang et al., 2024; Xu et al., 2025).

As a final note regarding persona, it may be helpful to recognize that in some instances the task is to act in a persona, and in other cases the optimization strategy for a task is to act in a persona.  For example, if the objective is concerned with writing culturally appropriate poems for a non-native culture, we may assign that cultural background to the system to help steer the behavior of the system, such as prompting it to "Act as an expert in culture, literacy and poetry from <location> and write a poem that …".  On the other hand, if the task objective is to act as a



customer or student with properties X so we may interact with you, in this case the persona is the central requirement which subsumes the overall task.

### Information Provisioning

Information provisioning strategies are concerned with the management of information provided to the system. This section addresses in context-example provisioning and information grounding via access and attention to additional artifacts, typically document, images or databases.

Following the machine learning literature, the LLM literature follows the practice of referring to examples in the data as "shots" with zero-shot meaning no examples, few-shot meaning a few examples and so on. Giving the system examples is a strong determinant of system behavior (Brown et al., 2020) as it makes explicit the pattern that the user is looking for, and the systems can often generalize beyond them (Garg et al., 2022; Zhou, J., et al., 2024). At the start of this decade these processes were not well understood and the mechanisms that allowed transformers to learn patterns not in the original training data was unknown. Abernethy et al. (2024) later showed it to be a property inherent in the attention mechanism.

While keeping in mind that any recommendation must be confirmed empirically and is task and system dependent, here we present 5 recommendations for maximizing ICL with examples.

**Recommendations for In-Context Example Prompting.**

*1. Use high-quality, representative examples.*

Examples should clearly represent the task the model is expected to perform. Effective demonstrations all define the same task, use similar inputs, produce the same kinds of outputs or data fields, and follow a consistent writing style. Min et al. (2022) showed that these structural



properties are the key drivers of in-context learning performance. Mixing examples that imply different task goals or output conventions, even if individually reasonable, consistently degrades accuracy.  Importantly, Wan et al. (2024) note the need to ensure system behavior is not dominated by "a few disproportionately influential examples"(p. 9).

In a manner analogous with humans, providing similar examples improves performance by aligning examples with the test instance (Rubin et al., 2022), while ensuring diversity among examples further boosts generalization (Yang et al., 2023). Selection strategies that best balance instance specificity and task coverage are nevertheless task- and model-dependent and may not transfer reliably across settings (Peng et al., 2024).

### *2. Standardize formatting and structure.*

Examples should use a single, clearly defined format for all instances and for the final query. Keep layout, field order, delimiters, section headers, and output style consistent, because even superficial formatting changes can produce large behavioral shifts (Min et al., 2022; Sclar et al., 2024) and format impacts vary across models (Voronov, et al., 2024).  While this may sound counterintuitive, Long et al. (2024) note that the fact that format markers, such as the symbols for parens or bolding, are often individually tokenized, thereby subjecting them to all the difficulties of prediction inherent in the transformer architecture.

### *3. Match the number of examples to the task.*

For straightforward tasks, a small number of clear examples, often on the order of a 1-5, can be sufficient, and additional examples may provide limited benefit. For more complex tasks—particularly those requiring reasoning or the suppression of default behaviors—using substantially more examples—ranging from dozens to hundreds (Brown et al., 2020), or even thousands (Agarwal et al., 2024)—when context length allows, can meaningfully improve



performance. At the same time, simply adding more examples without regard to quality, relevance, or consistency increases noise rather than improving results.

### 4. Account for example order effects.

The order in which examples appear can significantly influence model outputs, even when the set of examples is held constant. Examples placed later in the prompt often exert greater influence, which can unintentionally bias outputs if atypical or confusing cases appear near the end (Lu et al., 2022; Guo et al., 2024), an effect that varies across different models (Xiang et al., 2024).  Lu et al. (2022) demonstrated that reordering the same examples can shift performance from near chance to near state-of-the-art on the same task. Example order should therefore be treated as a design effect to be monitored and tested, rather than a neutral aspect of prompt construction.

### 5. Check robustness across prompt variations.

Given models' sensitivity to formatting, example selection, quantity, and ordering, performance should not be evaluated using a single prompt. Instead, assess multiple prompt variants and either select the most reliable configuration or aggregate results across prompts (Zhao et al., 2021; Sclar et al., 2024).  Evaluating a family of prompts provides a more realistic estimate of model performance and reduces the risk of over-interpreting unusually strong, or weak, results.  Zhuo et al. (2024) reported larger models were more robust to prompt variations when working with even modestly sized open source systems and Kim (2025) showed similar results on larger models.

**Information Grounding.**

Information grounding refers to prompt designs that require a model's responses to be supported by specific, identifiable source material rather than by the model's general background



knowledge. Grounding is most commonly used when working with external documents (such as PDFs) or with data retrieval-based systems using RAG, where the goal is to ensure that answers remain faithful to the provided texts and do not introduce unsupported information.

Recommendations are:

1. ***Focus the system's area of search to a specific range.***

When working with documents or retrieved material, prompts should instruct the system to rely only on the provided text and to indicate when the answer is not supported (e.g., "Use only the excerpts below; if the answer is not supported, respond 'not supported in the provided documents'"). Without this guidance, LLM-based systems often go beyond the supplied documents with unsupported or fabricated statements (Lewis et al., 2020; Gao et al., 2023; Li et al., 2024).

2. ***Help the system with structured textual boundaries.***

Passages relevant to the query should be shown directly to the system with clear labels, separators, or section markers and should be limited to a small, relevant subset rather than long or unfiltered documents (e.g., "Source A (Policy Manual §3.2): … Source B (Appendix C): …"). When dealing with complex situations, separate passage selection from answer writing (e.g., "Step 1: identify relevant passages; Step 2: answer using only those passages") as this decomposes the load on the system. (Gao et al., 2023; Peng et al., 2024; Jeong et al., 2024; Li et al., 2024).

3. ***Require explicit evidence references and allow 'not found' responses.***

Prompts should ask the system to connect each major claim to specific sentences, citations, or page references and should explicitly allow responses indicating that the information is missing (e.g., "For each claim, quote the supporting sentence; if none exists, state that



explicitly"). This keeps answers tied to the source material, exposes guesses that would otherwise sound plausible, and reduces false positives in tasks requiring factual accuracy (Bohnet et al., 2023; Saxena et al., 2024; Li et al., 2024).

### Chain-of-Thought and Process Elicitation

The finding that a simple cue such as *"let's think step by step"* could substantially improve reasoning performance (Kojima et al., 2022) demonstrated that prompting alone, without fine-tuning, can elicit multi-step reasoning from large language models that otherwise fail on complex problems. Additional work on chain-of-thought (CoT) prompting using explicit reasoning demonstrations further showed large gains on arithmetic, symbolic, and logical tasks (Wei et al., 2022). However, subsequent research has clarified important limitations of generic chain-of-thought prompting and motivates the following recommendations.

**1. Use explicit problem decomposition instead of generic "step-by-step" instructions.**

When tasks require multiple reasoning steps, prompts should explicitly break the problem into parts (e.g., "List the relevant variables, write the relationship between them, then solve") rather than relying on vague cues such as *"let's think step by step."* Early chain-of-thought work showed that even a minimal cue like *"let's think step by step"* can trigger multi-step reasoning in large language models (Kojima et al., 2022), and that providing worked reasoning demonstrations can substantially boost performance on hard reasoning benchmarks (Wei et al., 2022). But these approaches largely leave the *structure* of the reasoning to the model, which makes outcomes less stable than many practitioners expect.

Later studies show that clearly defined substeps—often implemented through simple planning stages, ordered subquestions, or light formatting—produce more reliable gains on complex tasks. Least-to-most prompting improves performance by decomposing problems into



simpler components and solving them sequentially (Zhou et al., 2022). Plan-and-solve prompting similarly finds that separating "plan" from "execute" reduces common failure modes in zero-shot chain-of-thought (Wang, L. et al., 2023). Empirical work also suggests that *what* the intermediate steps contain matters; it's not just "having steps," but having the *right* intermediate structure (Wang, B., et al., 2023).

**2. Request explicit reasoning only when intermediate steps are necessary.**

Prompts should ask the system to show its reasoning only when the task genuinely depends on intermediate calculations or logical dependencies. Chain-of-thought prompting is consistently useful on math, symbolic, and formal reasoning tasks (Wei et al., 2022), but it can be unnecessary—or counterproductive—when the task is better solved by pattern recognition or implicit learning.

Recent evidence is fairly direct here: in controlled experiments on implicit statistical learning tasks, requiring step-by-step reasoning reduced accuracy, including large drops (up to ~36 percentage points in some conditions) because "thinking aloud" disrupted simpler and more effective solution paths (Liu, R., et al., 2025). Accordingly it is best to treat explicit reasoning as a tool you turn on when you need intermediate computation, not a default setting.

**3. Be cautious when using chain-of-thought with reasoning-optimized models.**

For models explicitly trained for internal multi-step reasoning (e.g., OpenAI's o-series reasoning models), prompting for full step-by-step explanations should be used sparingly. In practice, these models already carry out multi-step reasoning internally, and OpenAI's guidance emphasizes that requesting step-by-step explanations is often unnecessary and may not improve results (OpenAI, n.d.b). The better pattern is usually to specify constraints, ask for a clear final answer, and (when appropriate) request verification checks rather than a full reasoning trace.



This caution also lines up with research in planning-style domains: chain-of-thought can look persuasive while being brittle or misleading, and its usefulness can depend heavily on prompt phrasing and the task setup (Stechly et al., 2024). For reasoning-optimized models, you typically get more leverage from *problem specification and validation* than from consistently forcing long, externalized chains.

### System Interaction Management

System interaction management concerns how users engage with language models across turns, rather than how a single prompt is phrased. A common and consequential misuse pattern is treating LLMs as static information retrieval tools, issuing single queries and expecting optimal answers, rather than as interactive systems designed for clarification, feedback, and refinement. This interaction style which this author sometimes calls, "Google brain" reflects the fact that many of us have long histories of specific human-computer interaction styles that do not align with interactional styles of a chatbot interface. The recommendations below focus on interaction level practices that align with how these systems are trained and how they behave.

**1. Treat model use as an iterative interaction, not a one shot query.**

LLMs are optimized for conversational turn taking and benefit from incremental refinement rather than fully specified initial prompts. Instead of assuming a system can solve a task in one step, users will often need to clarify goals and respond to intermediate outputs, for example asking "Before revising this, tell me what is underspecified," followed by "Now revise section 3 using those clarifications." This interaction pattern surfaces ambiguities early and reduces downstream misalignment. Common models are trained with instruction following (via SFT) and respond particularly well to iterative guidance (Wang, Y., et al., 2023).

**2. Actively manage conversational context.**



Although models condition on prior turns, they do not reliably retain or prioritize all earlier information. Accordingly, it is helpful to periodically restate or foreground important constraints, definitions, or decisions such as "Summarize the assumptions we have agreed on so far," or "For the rest of this task, prioritize the requirements in the last paragraph." Empirical work on long context behavior shows that information in extended prompts is often underutilized, especially when it appears mid context (Liu, N. F., et al., 2024). While these effects are likely somewhat ameliorated by increases in context window size, user expectations and additional demands may counter that effect.

**3. Use structure to coordinate interaction, not just to format outputs.**

Structural cues such as delimiters, labels, and templates are most effective when they clarify how different parts of the interaction should be interpreted. Separating background, task instructions, and constraints, for example "<context>…</context> <task>Analyze…</task> <constraints>Limit to 250 words</constraints>," reduces ambiguity about what information the system should rely on. Formatting and separator choices materially affect how models parse and weight inputs, leading to measurable performance differences (Park, Y., et al., 2024; He, J., et al., 2024). Providing structured language for both inputs and outputs helps the systems take advantage of the "privileged properties of code" discussed above.

**4. Use the system to help design, refine, and distill effective prompts.**

Users should consider asking the system for help crafting prompts, especially for complex or unfamiliar tasks, for example "Tell me how to ask you this question to get a technical answer." Likewise, after a long interaction has converged on a satisfactory result, it is often useful to ask the system to distill the exchange into a single reusable prompt, for example "Write a prompt that would produce this outcome without the full dialogue." Research on self



instruction and prompt optimization shows that models can abstract effective prompt structures from interaction history, improving efficiency and reproducibility in later use (Wang, Y., et al., 2023; Xiang, J., et al., 2025).

**Prompt Engineering (PE)**

We can think of the prompting best practices discussed above as "knobs" that help steer the LLM behavior imperfectly but directionally.  Prompting may be used in playful or highly exploratory modes or in clearly objective driven contexts.  When we need a coherent approach for the latter, we consider that work PE as illustrated in Figure 14.



# Prompt Engineering Workflow

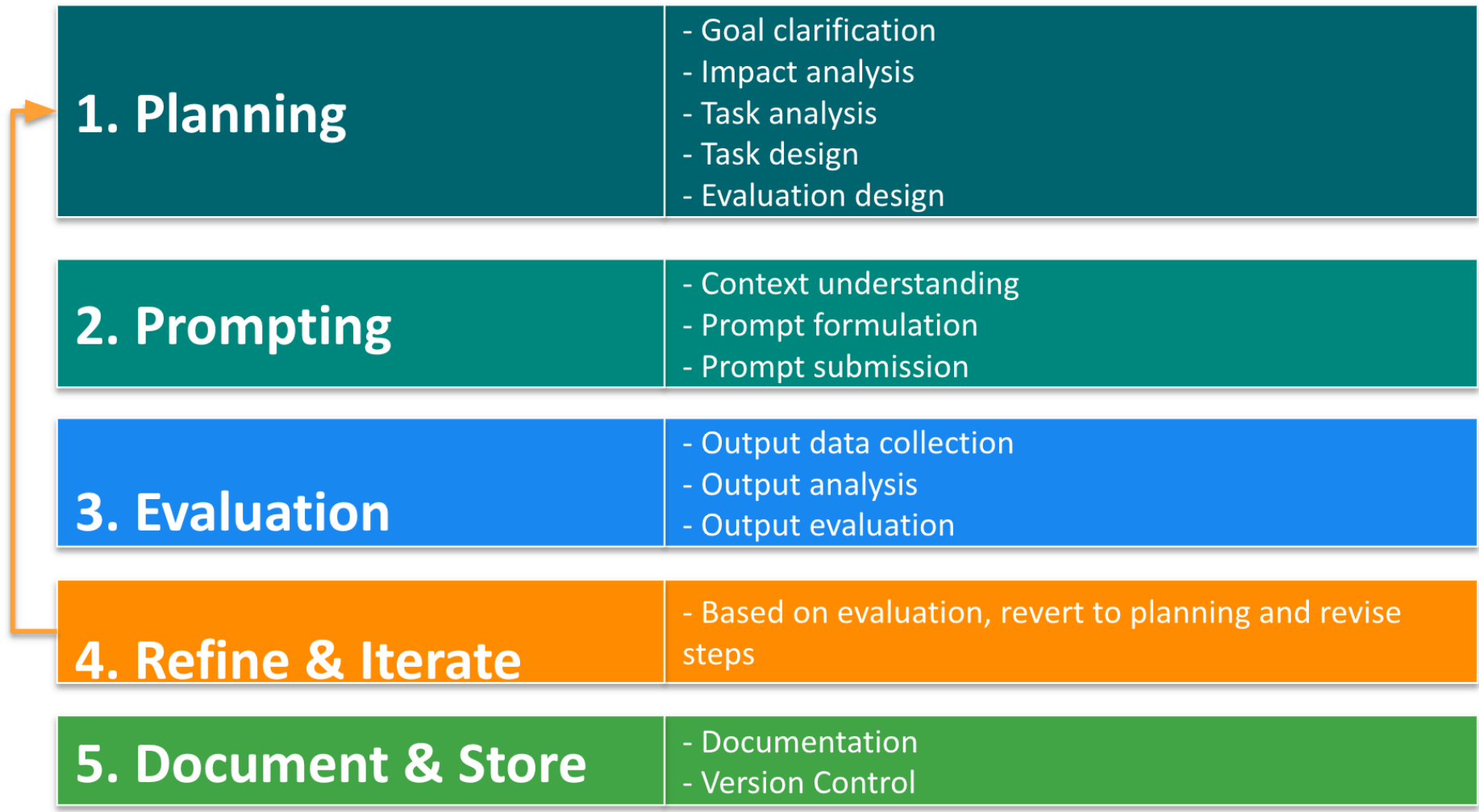

Figure 14. Overview of the Prompt Engineering Workflow



As the reader can see, PE is understood as an empirical and cyclical process in which prompting is a necessary but insufficient element.  Here, the types of activities most aligned with these phases are briefly discussed.

*Planning*

In this framing, the five-step process starts with clarifying the goal of the endeavor: Is a formal process necessary? This would likely depend on whether there are any potential impacts that would have negative impacts that would need to be mollified, or if the process was going to be repeated frequently, say with a large number of students or faculty.  In the goal formulation phase one might determine that GAI is not even the appropriate method for the goals being formulated. Impact analysis would consist of an effort to understand costs along monetary as well as personal, social, and environmental dimensions.

Assuming the goal is clear and the costs are acceptable, the understanding of the tasks to be performed must be specified, as well as the method for evaluating the suitability of the output in alignment with the goals.  Task design in alignment with project goals is similar to assessment design and our experience is that LLM evaluation and prompt engineering benefit greatly from the application of Evidence Centered Design (ECD; Behrens et al., 2004; Mislevy et al., 2012). Behrens (2025b) provides open access to an extended guide on how to apply ECD for the creation of custom LLM assessments in this context (Orchard et al., 2025) and Liu, Y., et al. (2024) provide an ECD framework for developing AI benchmarking assessments (BECD).

*Prompting*

The topic of prompting is not addressed here as it has been discussed at length above.



*Evaluation*

### *Benchmarks.*

As with task and overall assessment design, system evaluation can be conceptualized as an educational assessment endeavor in which the performance of the system on a range of tasks is used to judge its performance.  In the AI literature, benchmarks or "evals" as they are sometimes called, represent a cottage industry similar to test construction.  Ni et al. (2025), for example, enumerate 283 different benchmark item sets across general capabilities (knowledge, linguistic core, reasoning), domain specific benchmarks (natural sciences, humanities & social sciences, engineering & technology) and target specific benchmarks (risk & reliability, agent behavior, others).  In large scale settings, evaluation will happen with statistical analysis on repeated samples of LLM performance.  Behrens (2025b) provides an open resource of annotated links to approximately 50 common benchmarks with detailed examples for instructional use.  The vast majority of benchmark data sets are committed to the open data repositories at http://huggingface.co.

### *Simulation Tools.*

The probabilistic nature of LLM behavior raises the question of "what would happen if I gave the same prompt - task - system combination again?", and what would happen in the long run?  For researchers that are interested in running experiments investigating LLM behavior against specified tasks without any computer code, we suggest the reader explore the Chainforge software (Arawjo, 2024) that is accessible at http://chainfoge.co.  The system uses no-code drag and drop interfaces that allow configuration of numerous experimental configurations that support LLM-based research and instruction.



Figure 15 is a screen shot of a Chainforge flow.  Data moves from left to right from a text field node that contains benchmark tasks (aka test items) and is sent to a number of LLM systems via API keys followed by an interactive chatbot widget programmed to respond 1 of 4 different ways.  Results are scored by the LLM scorer nodes in the center of the display, and the results are passed on to visualization nodes.  Different configurations are obtained by connecting lines between nodes and filling out text boxes or pull down menus associated with the nodes. This provides an easy and scalable way to conduct experiments on prompt, model and task effects at scale. Data can be exported and system configurations can be saved as descriptive JSON files.

Among the many values of such a system is that it allows systematic evaluation of the long run behavior of a system.   Users are not required to have advanced coding skills to investigate such questions at scale.  Orchard et al. (2025) report that the author makes substantial use of Chainforge in their class including lessons regarding experimental design, data collection and LLM analysis.

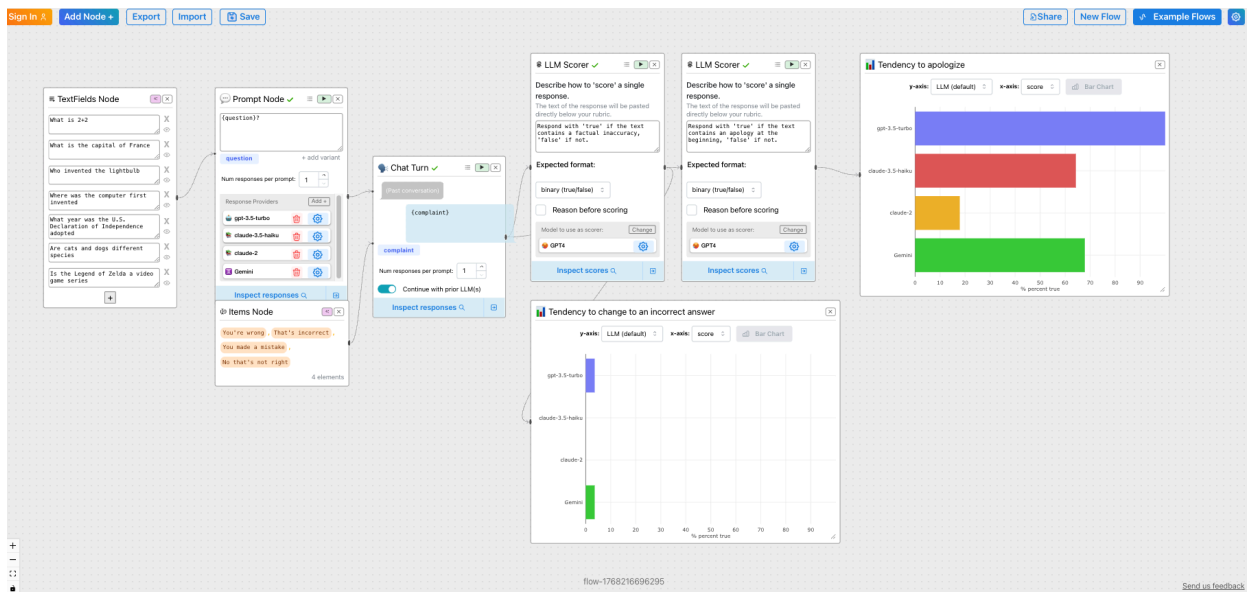

Figure 15. Screenshot of Chainforge Experimental Setup



### *Statistical Analysis.*

The statistical analysis of results from LLM behavior is an area of potential research for educators and psychometricians given its parallels to human proficiency evaluation. For example, Miller (2024) criticized practices seen in the LLM research community including failing to account for hierarchical nesting and other dependency effects (cf. Gallo et al., 2024) as well as failing to consider statistical power. Miller also argued against lowering temperature for variance reduction purposes, as it changes the system being measured and can actually increase overall estimation variance by inflating true score variability. Importantly, he also recommended, when feasible, that researchers do not report the LLM behavior on fixed-response tasks, but rather the output probabilities which are more akin to latent proficiency scores than observable scores.

In other work applying assessment theory to LLMs, Degen (2025) provided a framework laying a range of concerns in AI research that could benefit from application of Generalizability Theory. Other authors have applied Item Response Theory to LLM analytics (Polo et al., 2024) though some have suggested limitations of this approach (Madaan et al., 2024). As Dobriban (2025) notes "A comprehensive statistical framework for the evaluation of generative AI systems is yet to be developed"(p. 19). Abdurahman et al. (2024a) provides an important set of methodological cautions as a counterpoint to unbridled enthusiasm in this domain.

### *Refine and Iterate*

Measurement and evaluation of model misfit is a fundamental operation in both quantitative and qualitative research methods (Behrens & Smith, 1997). Qualitatively it means attempting to determine the root cause of the behavior and modify and submit an additional prompt to try to steer the system toward the goal. When the prompting will occur repeatedly, for



example every time a user logs into learning software for a particular task, automatic prompt engineering techniques (APE) are increasingly more likely to be used. In these systems, the success or failure signals from the system's performance on tasks are used iteratively to test a large number of prompting combinations until the optimal prompt is found. For example, DSPy (Khattab et al., 2023) is an open source Python library that supports the creation of automated prompting and inference optimization pipelines that searches prompt variations until an optimal configuration is found. Where prompt implementation is repetitive at large scale, this is an increasingly popular approach dubbed "let the model write the prompt". See Chang et al. (2024) for additional detail on APE and some interesting historical details on prompting evolution.

### *Document & Store*

The final step in our PE process is simply good business and good housekeeping: Keeping track of what we have learned and how best to re-use our knowledge in the future.

### *Prompting Summary*

While the format of the book requires these methods be presented individually, in practice they work in an integrated and additive manner. For example, Suzgun & Kalai (2024) created an automated meta-prompting system that broke problems down into sub-tasks (problem decomposition / CoT), assigned appropriate persona experts for each sub-task (Identity specification), and applied self-reflection & monitoring across the overall activities (orchestration). They found this method to significantly enhance performance over standard and automated-expert assignment across a range of task types.

When individuals work directly with GAI systems (rather than in an automated manner), these practices likewise reflect a shift from old information-retrieval paradigms of one-time inquiries to actively managed iterative interactions, with iteratively shaped context, constraints,



and goals over time.  As noted at the outset of the chapter, dealing with iterative and

non-deterministic behavior is the unusual, but now standard, requirement for users of generative

systems.

## Remaining Themes

This chapter, focused on the fundamental logics, techniques and technologies for GAI

and eschewed many of the more recent developments.  Nevertheless, we have seen hints of them

here and there through the literature.  Undoubtedly the broad understanding of the landscape is

the most important skill  going forward, insofar as specific details will change over time.

Nevertheless here are 2 trajectories the reader should consider as they represent active and

dynamic shifts already underway.

### Agents

As discussed elsewhere in this volume, agentic workflows are a natural next step in the

development of LLMs in computing.  An agentic workflow is a computing environment in which

LLMs are called upon to perform specific cognitive tasks that may involve complexities

including tool use (e.g open a web browser, search for an item and purchase it) and group-agent

collaborations.  This is a sub-domain in which there are numerous unresolved issues such as

optimal prompts and patterns for agentic collaboration, but also high and immediate value in

many current educational use cases.  The reader is encouraged to understand the privacy and

security risks for themselves and their educational stakeholders as these new technologies

evolve.

### Self Empowering Code Use

As noted above, the linguistic form of computer programming,  and highly structured file

systems or markup languages, hold a privileged place in the GAI ecosystem and this trend is



accelerating. The reader can expect to see continued focus on improvements in coding tools with dramatic amounts of funding and energy aligned with this need.

In addition, we should expect to see cyclical synergies between code use and code production. Consider, for example, if we wanted to teach an LLM how to design and create Chainforge experiments, described above. To accomplish this we would look for data that could be used to train it by giving it lots of examples per out prompting strategies. One way to do this is to find lots of example Chainforge files on the internet and feed them to an LLM. This is often done by sharing online Github code repositories.

After a certain number of examples have been made accessible to the LLM, it would learn the code scheme that defines the system and eventually be able to produce files to requirements, at least as strong first drafts. If sufficient quality was obtained, then it is likely the system could generate its own files that it could in turn use to re-train itself, potentially with custom examples and goals. Agentic workflow systems are typically also represented in JSON or other coding patterns and likewise could be learned by LLMs to create their own workflow patterns. In sum, we are moving towards a world of dynamic AI self training in which highly structured languages and environments most benefit.

## Conclusion

We began this chapter with the metaphor of subjects in a great experiment, confronting an apparatus that neither participants nor experimenters fully planned nor understand. Having traced the foundational elements of GAISs and what we know about best practices for working with them, it is hoped that the reader has a stronger understanding about how these systems work and is better equipped to engage with that apparatus and hopefully, each other.



Among the themes that emerges from this overview is a fundamental tension at the heart of GAI use: a conflict between user expectations shaped by decades of deterministic computing and the fundamentally probabilistic nature of large language models. When we approach a traditional software application, we carry implicit assumptions that the system will behave consistently, that its behavior traces back to specific intentional design, and that the boundaries of its use are well determined. Generative AI systems violate these assumptions with behavior that varies across identical prompts and emerges from opaque parameter interactions rather than explicit programming. This mismatch undergirds much of the confusion, frustration, and anthropomorphization that surrounds these technologies. It also explains why so many recommendations from one context fail to generalize to another. Our predictions of its predictions are always falling short in surprising ways.

For educational researchers, this paradigm shift arrives with additional complications. The B2C commercialization strategy that brought ChatGPT and its successors directly to consumers, especially students, has created conditions of expertise inversion seldom seen in educational technology. Echoing the history of the internet and mobile phones, students have, in many cases, accumulated interactions with these systems than their instructors, developing intuitions and workflows that faculty may not share or even recognize. This inversion challenges traditional knowledge hierarchies and complicates efforts to develop institutional policies, pedagogical frameworks, or research protocols. Researchers must contend not only with the technical complexity of the systems themselves but with the social fact that their participants may hold different, sometimes more sophisticated, mental models of how these tools function.

The rapid evolution of these systems compounds these challenges. Prompts that were effective two years ago may fail today; capabilities that required fine-tuning are now addressable



through simple instruction; systems marketed as "language models" have quietly become multimodal, processing images and video alongside text. Researchers who treat generative AI as a stable apparatus risk discovering that their experimental conditions have shifted mid-study. Those who depend on commercial products cede some control to vendors who may update, modify, or deprecate features without notice—as the sycophancy episode with GPT-4 demonstrated. In light of this continuous change, researchers must now add to their work a long list of professional development requirements, to keep up with the ever changing dynamics. At the same time, research methods involving these systems also evolve. For example researchers using these systems must document not only the model used but the version, the interface, the date of interaction, and where possible, the system prompt and parameter settings. Moving forward, such documentation is not merely good practice; it is a precondition for interpretability and replication.

Educational researchers need not shrink from unfamiliarity with advanced computing. If the study of GAISs is, as Kambhampati suggested, "a kind of natural science, even if an ersatz one," then educational researchers trained in empirical methods, familiar with measurement challenges, attentive to validity threats, are well positioned to contribute. The same skills that allow us to evaluate the reliability and validity of assessments, to design studies that account for confounds, and to interpret hierarchical effects, apply directly to the evaluation of AI system behavior. Evidence Centered Design provides structure for specifying what we want systems to do and what evidence would demonstrate success, while psychometric frameworks like Generalizability Theory offer tools for partitioning variance in LLM outputs. The challenges are new, but they are not wholly unfamiliar.



Amidst all this change in our core professional activities, we are likewise embedded in a larger social world that is rapidly changing as well. In just the last few months at the end of 2025 we are seeing dramatic improvements in these GAISs ability to code complex software systems with comparatively little direction. This will lower the cost of entry to students, faculty and entrepreneurs creating software while also leading to the need for rapidly updated curriculum and unavoidable job loss. The pace of technological change is likely to outrun the social, economic, and political systems currently in place for the foreseeable future.

Finally, vigilance must extend beyond technical evaluation to the social and ethical issues these systems introduce. The biases embedded in training data, the potential for representational harm, the risks of model collapse and diversity loss, the unresolved questions of copyright and intellectual property, are not peripheral concerns to be addressed after the "real" work of technical application is complete. They are integral issues concerning what these systems are and how our interactions with them will shape educational research and practice. Researchers bear responsibility not only for documenting what works but for attending to what these systems might erode, exclude, or distort.

The apparatus will continue to evolve, likely in directions and speeds we cannot anticipate. Nevertheless, the experiment has been thrust upon us and we are left to do our best to cultivate a stance of informed engagement. As the experiment continues, we remain both its subjects and, increasingly, its interpreters, building on a long line of existing research and the educators' values of putting the students' overall well being first.

**\*\*\* End of Document (Preprint only)**